\documentclass[aps,twocolumn,superscriptaddress]{revtex4-1}
\usepackage{graphicx,amsmath,hyperref, xcolor}
\usepackage{amssymb}
\usepackage[utf8]{inputenc}
\usepackage{multirow, makecell, comment} 
\usepackage{braket}
\usepackage{lineno}
\usepackage{lipsum}
\newcommand{\fla}[1]{\begin{flalign}#1\end{flalign}}
\def\mean#1{\langle\,#1\,\rangle}
\definecolor{Gray}{gray}{0.93}
\definecolor{green1}{RGB}{0, 161, 73}
\definecolor{blue1}{RGB}{91, 155, 213}
\definecolor{red1}{RGB}{237, 70, 70}
\usepackage[colorinlistoftodos]{todonotes}

\newcommand{\beq}{\begin{equation}}
\newcommand{\eeq}{\end{equation}}
\newcommand{\ba}{\begin{eqnarray}}
\newcommand{\ea}{\end{eqnarray}}
\newcommand{\bea}{\begin{eqnarray}}
\newcommand{\eea}{\end{eqnarray}}
\newcommand{\bma}{\begin{subequations}}
\newcommand{\ema}{\end{subequations}}
\newcommand{\bwt}{\begin{widetext}}
\newcommand{\ewt}{\end{widetext}}
\def\abs#1{|\,#1\,|}

\def\mean#1{\langle \,#1\, \rangle}
\begin{document}
\title{Broadband quantum memory in a cavity via zero spectral dispersion}
\author{E. S. Moiseev}
\affiliation{Institute for Quantum Science and Technology, University of Calgary, Calgary AB T2N~1N4, Canada}
\author{Arina Tashchilina}
\affiliation{Department of Physics, University of Alberta, Edmonton AB T6G~2E1, Canada }
\affiliation{Institute for Quantum Science and Technology, University of Calgary, Calgary AB T2N~1N4, Canada}
\author{S. A. Moiseev}
\affiliation {Kazan Quantum Center, Kazan National Research Technical University named after A. N. Tupolev,  Kazan, Russia}
\author{Barry C. Sanders}
\affiliation{Institute for Quantum Science and Technology, University of Calgary, Calgary AB T2N~1N4, Canada}

\begin{abstract}
We seek to design experimentally feasible broadband, temporally multiplexed optical quantum memory with near-term applications to telecom bands.
Specifically, we devise dispersion compensation for an impedance-matched narrow-band quantum memory by exploiting Raman processes over two three-level atomic subensembles, 
one for memory and the other for dispersion compensation.
Dispersion compensation provides impedance matching over more than a full cavity linewidth. Combined with one second spin-coherence lifetime the memory could be capable of power efficiency exceeding 90\% leading to $10^6$ modes for temporal multiplexing.
Our design could lead to significant multiplexing enhancement for quantum repeaters to be used for telecom quantum networks.
\end{abstract}
\maketitle
\section{Introduction}
Optical quantum memory provides on-demand storage for quantum information encoded into light and furthermore provides on-demand release of this quantum information~\cite{QMReview, Hammerer2010,Khabat2016, CHANELIERE2018},
which is vital for scalable quantum communication~\cite{Razavi2009}.
As optical quantum memory continues to be imperfect,
performance of the quantum memory is quantified by relevant task-related figures of merit such as power efficiency~$\wp$
(ratio of output to input power),
time-bandwidth product~$\eta$
(ratio of maximum permitted pulse delay to pulse duration) and storage time~$\tau$ (usually defined as the time required for a 50\% drop in power efficiency).
Good optical quantum memory
would demonstrate an 
efficiency of $\wp=0.75$,
$\tau=100$~ms~\cite{Razavi2009}
and $\eta=10^3$~\cite{Simon2007}
with a communication rate exceeding hundreds of megahertz,
which implies pulse duration no longer than a few nanoseconds.
To date,
no theoretical proposal for experimental implementation comes close to satisfying these three criteria simultaneously.
Here we present our design for a feasible broadband quantum memory
which employs three-level atoms coupled to an optical resonator,
with atomic density, cavity quality, and Raman-laser power and detuning chosen such that inverse cavity lifetime~$\kappa$
is proportional to effective collective coupling constant,
$\eta>10^6$, $\wp>0.9$,
and $\tau>1$~s,
leading to $10^6$ modes for multiplexing by increasing the memory bandwidth by more than $\kappa$.

To date several protocols can theoretically provide multimode memory fulfilling the targets $\wp$, $\eta$ and~$\tau$. 
One category of proposals for multiplexed broadband quantum memory suggests an atomic frequency comb with two-level atoms but with $\tau<100~\mu$s~\cite{AfzeliusPRA2010};
however, storage time can be extended by using pulse control 
albeit at the cost of introducing significant noise due to pulse-area fluctuations, especially in the optical domain.
Another multiplexed broadband optical quantum memory proposal is revival of silent echo (ROSE) that also suffers from noisy $\pi$-pulse control~\cite{Dajczgewand:14}.
The proposal for impedance-matched off-Resonant echo memory is closest to the target $\wp$ and~$\tau$, but typically small optical depth of off-Resonant transition limits the bandwidth to sub-MHz~\cite{SMoiseev2013}.


Amongst the many optical quantum memory experiments~\cite{QMReview},
we highlight the four most relevant based on being broadband and having achieved $\eta>10^3$.
One of these experiments achieves $\wp>0.92$ but low $\tau\sim200~\mu$s~\cite{Hsiao2018}.
Another experiment achieves moderate 
retrieval efficiency $\wp=0.76$ but long $\tau=220$~ms~
\cite{Yang2016}.
A third experiment of interest demonstrates Telecom O band memory with a low $\wp=0.07$ but a good $\tau=100$~ms~\cite{Radnaev2010}.
The three experiments mentioned so far are all single mode but a fourth experiment demonstrates Telecom~C band memory, unfortunately for negligible~$\wp$ and with $\tau=10~\mu$s~\cite{Craiciu2019}.
The second and fourth experiments in this list have been performed in cavities to enhance atom-field coupling with the second experiment performed in a narrowband cavity and the fourth experiment in a broadband cavity with the latter suffering from impedance mismatch and intrinsic cavity losses.

We regard incorporation of a cavity as vital for optical quantum memory to achieve requisite targets for~$\wp$, $\eta$ and~$\tau$
while being multimode.
Optical quantum memory requires efficient transfer of quantum information from photons into long-lived excitations of matter,
which can not be achieved by a weak coherent light-atom interaction.
The straightforward solution is to increase the number of atoms in an ensemble,
which would involve long cells for hot gases
or dense traps for cold gases~\cite{Sparkes2013} or special a sub-wavelength arrangement of atoms~\cite{ManzoniNJP2018}.
Alternatively optical resonators can be used
to enhance light-atom interaction
but at the expense of narrowing the bandwidth~\cite{GorshkovPRA12007}
because weak coupling demands a high-quality optical cavity, 
which leads to proportional narrowing of the quantum-memory bandwidth~\cite{Moiseev2010b,SabooniPRL2013}. 

White-light cavities were first introduced as an optical cavity with double~$\Lambda$ atoms inside with the purpose of combining the benefit of a broadband cavity with a large quality factor~\cite{WICHT1997431}.
These double~$\Lambda$ atoms serve to compensate deleterious dispersion.
White-light cavities have been proposed for increasing the bandwidth-sensitivity product for gravitational wave detection~\cite{WICHT1997431}
and linear~\cite{Xu2012} and non-linear~\cite{Li2016} optical switching.
Experimentally,
white-light cavities were demonstrated with bandwidth increase ranging from from 4- to almost 10-fold with warm rubidium vapour placed in a ring cavity~\cite{Pati2007, Wu2008}.
Further study showed that quantum noise generated by atoms ruins enhancement of the shot-noise-limited sensitivity~\cite{Ma2015}.

\begin{figure}
\includegraphics[width=0.98\columnwidth]{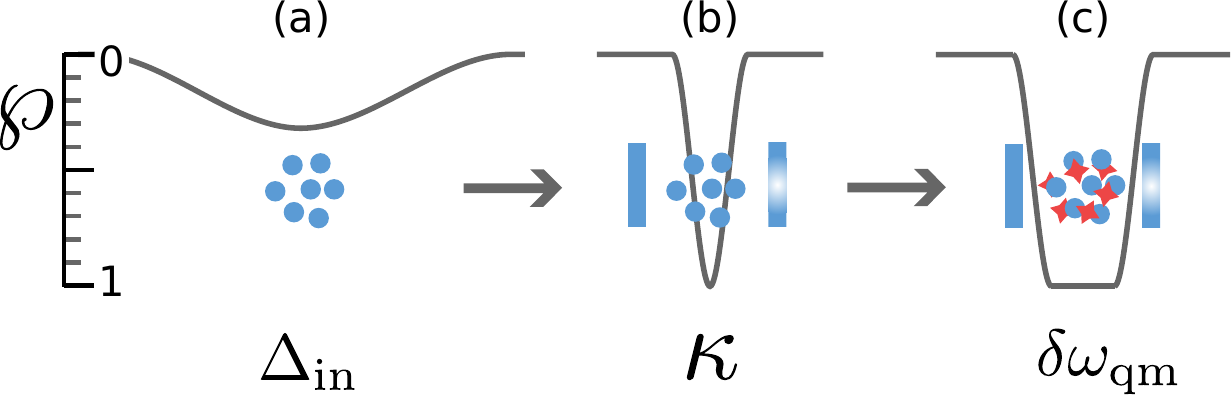}
\caption{%
How absorption of the medium changes for different protocols: (a) for a free-space ensemble, absorption is dictated by optical depth and by inhomogeneous broadening; (b) if the ensemble is placed into a resonator under impedance-matched conditions, one can observe 100\% absorption within $\sim\kappa/3$; (c) we add a second ensemble, which could extend the highly absorptive width beyond $\kappa$.}
\label{fig:idea}
\end{figure}

We build on a proposal for cavity-enhanced optical quantum memory~\cite{SMoiseev2013},
which has the advantage of optimal atom-field interaction thereby enhancing field mapping into long-lived states but the disadvantage of being narrowband.
Our new design combines dispersion compensation with optical quantum memory, with each of the two requisite media being subensembles of the atoms in the cavity.
Whereas white-light cavities were proposed to have double-$\Lambda$ atoms, our concept involves regular $\Lambda$-type atoms,
but inhomogeneous broadening results in subensembles that can be selected by hole-burning methods~\cite{Nilsson2004}.
One subensemble serves as optical quantum memory and another subensemble serves as a dispersion compensator as  depicted in Fig.~\ref{fig:idea}.

Next we present our quantum treatment for storing the incoming field in Raman echo quantum memory  operating in an impedance-matched white cavity regime and show how it extends quantum memory bandwidth. 
At the end we analyze the noise induced by the imaginary part of the abnormal dispersion and present a regime of a noise free operation.
Finally, we discuss possible experimental implementations. 

\section{Mathematical description}
\begin{figure}
\includegraphics[width=0.98\columnwidth]{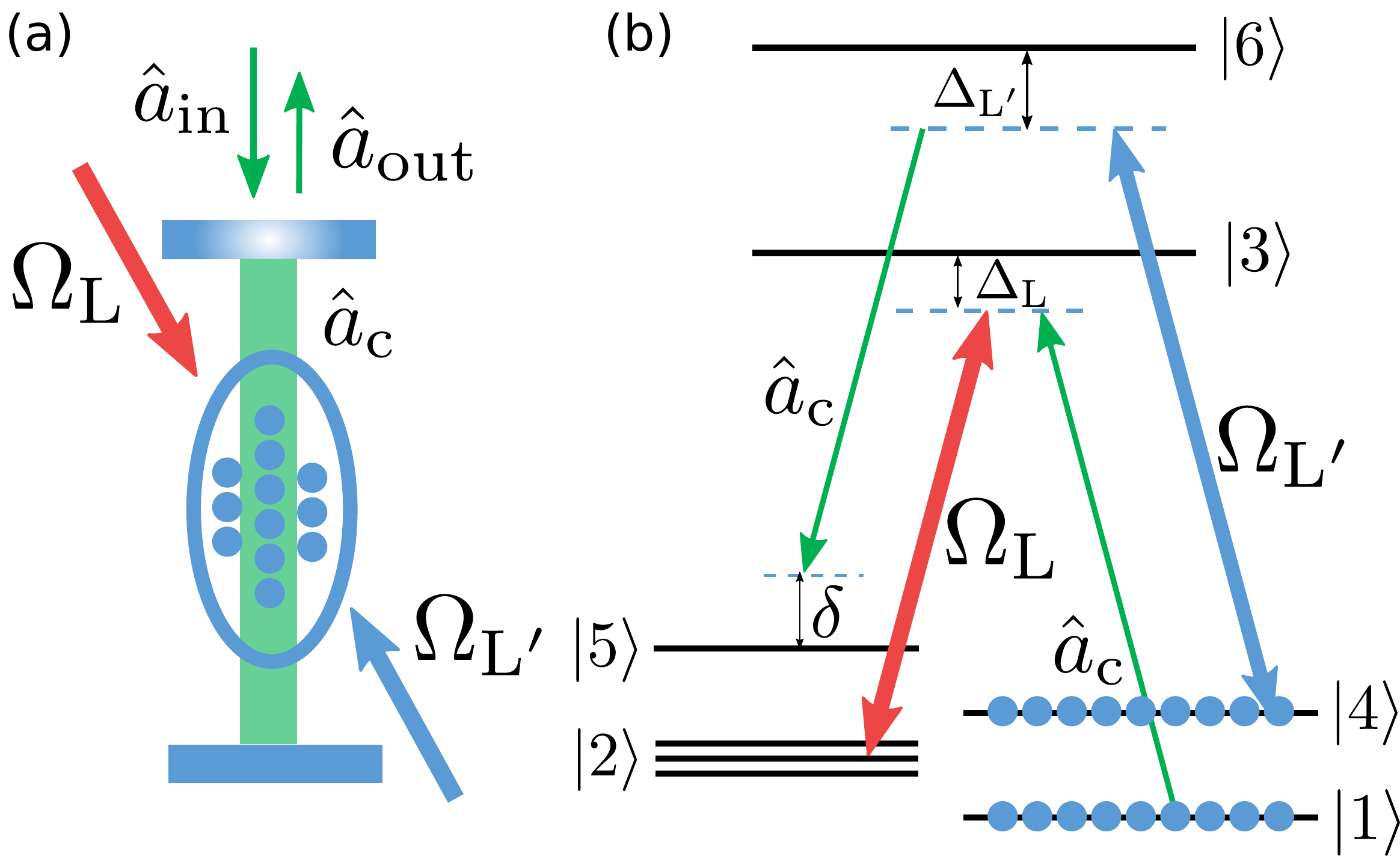}\quad
\caption{%
(a)~The atomic medium is coupled to a Fabry-Perot cavity.
(b)~Atomic levels and quantum transitions: strong control laser field $\Omega_L$ provides an absorption for optical cavity field $\hat{a}$ via two-photon Raman transition $\ket1\rightarrow\ket3\rightarrow\ket2$; the dispersion compensation (DC) laser field $\Omega_{\text{L}'}$ through the Raman scattering on second lambda scheme $\ket4\rightarrow\ket6\to
\ket5$ realizes dispersion compensation mechanism for extending impedance-matched cavity linewidth.}
\label{fig:2lambda}
\end{figure}

Now we describe our scheme.
Specifically, we propose that $\Lambda$-type atoms are placed inside a high-quality single-mode cavity depicted in Fig.~\ref{fig:2lambda}(a).
The atomic level structure is depicted in Fig.~\ref{fig:2lambda}(b).
By suitable optical pumping~\cite{amari2010towards},
ground atomic levels $\ket1$ and $\ket4$ are initially populated, 
whereas the other ground levels $\ket2$ and $\ket5$ are unpopulated.
Our level~$\ket2$
is represented by a manifold,
reflecting controllable inhomogeneous broadening $\delta^j_{21}$ to enable on-demand introduction of photon echo.
Two lasers~L and~L$'$ drive the atoms and are
quasi-continuous beams with wave vectors~$\vec{k}_\text{L}$ and~$\vec{k}_{\text{L}'}$ and Rabi frequencies~$\Omega_\text{L}$ and~$\Omega_{\text{L}'}$.
These two beams are detuned by~$\Delta_\text{L}$ and~$\Delta_{\text{L}'}$
from the $\ket2\leftrightarrow\ket3$
and $\ket4\leftrightarrow\ket6$ transitions, respectively.

The optical state to be stored enters the cavity at the input port,
and the annihilation operator
for this input field is denoted~$\hat{a}_\text{in}$.
The output-field annihilation operator is~$\hat{a}_\text{out}$,
and, within the cavity,
the corresponding cavity-field annihilation operator is ~\cite{Gardiner1985}
\begin{equation}
\label{eq:acaoutain}
    \hat{a}_\text{c}
        =(\hat{a}_\text{out}
            +\hat{a}_\text{in})
                /\sqrt{\kappa},
\end{equation}
which couples transitions $\ket1\leftrightarrow\ket3$
and~$\ket5\leftrightarrow\ket6$
with single-photon detunings~$\Delta_{\text{L},\text{L}'}$
and two-photon detuning~$\delta$, respectfully.

We now describe mathematically the interaction between the cavity field and the atomic transitions
in the linear response regime,
which is valid for a weak cavity field,
hence low population outside level~$\ket1$
and near-unity population in level~$\ket1$.
We obtain the  system of equations for slowly varying operators of the cavity mode
$\hat{a}_\text{c}$
and atomic coherences for each $j^\text{th}$ atom 
defined by
$\hat{P}^j_{pq}
    :=\ket{p}^j\!\bra{q}$
for $p,q\in\{1,\ldots,6\}$.
Operators~$\hat{P}^j_{pp}$
yield population in level~$\ket{p}$
with mean~$\bar{P}^j_p:=\left\langle\hat{P}^j_{pp}\right\rangle$.

We let~$\kappa$
denote the coupling constant for the cavity mode with external field modes,
and~$\vec{k}_\text{c}$ is the wave vector of the cavity mode.
Coupling constants~$g_{13}^j$ and $g_{56}^j$ are for cavity-mode interaction with atomic transitions $\ket1\leftrightarrow\ket3$ and $\ket5\leftrightarrow\ket6$.
We employ~$\{\hat{F}^j_{pq}(t)=\hat{F}^{j\dagger}_{qp}(t)\}$
to represent generalized $\delta$-correlated Langevin noise~\cite{Scully1997} with correlators~\cite{FleischhauerPRA2013, lauk2016quantum} 
\begin{equation}
\label{eq:corr}
    \mean{\hat{F}^j_{pq}(t)\hat{F}^j_{qp}(t')}=D^j_{pp}(t)\delta(t-t'),\,
    D^j_{pp}\propto \gamma_{pq} \bar{P}^j_p
\end{equation}
with~$D$ representing diffusion coefficient,
which is proportional to mean population.
Physically,
$\hat{F}^j_{12}$ and~$\hat{F}^j_{13}$ represent spontaneous emission.
Then the three dynamical equations are~\cite{GorshkovPRA2007, SMoiseev2013}
\fla{
\dot{\hat{a}}_\text{c} =& -\frac{\kappa}{2} \hat{a}_\text{c}- \text{i} \left( \sum^{N_1}_{j=1} g_{13}^j \hat{P}^j_{13}+ \sum^{N_2}_{j=1} g_{56}^j \hat{P}^j_{56}  \right)\text{e}^{-\text{i}\vec{k}_\text{c} \cdot\vec{r}_j} \nonumber\\
&+ \sqrt{\kappa} \hat{a}_\text {in},
\label{eq:a_c}\\
\dot{\hat{P}}^j_{13} =& -(\gamma_{13}+\text{i}\Delta_{\text{L}})\hat{P}^j_{13}
-i g_{13}^{j*}\text{e}^{\text{i}\vec{k}_\text{c}\cdot \vec{r}_j} \hat{a}_\text{c} \bar{P}^j_1 
    \nonumber\\&
- i\Omega_\text{L}\text{e}^{\text{i}\vec{k}_\text{L} \cdot\vec{r}_j} \hat{P}^j_{12}+\hat{F}^j_{13}, 
\label{eq:P_13}\\
\dot{\hat{P}}^j_{12} =& -(\gamma_{12}+\text{i}\delta_{21}^j ) \hat{P}^j_{12} - i\Omega_\text{L}^*\text{e}^{-\text{i}\vec{k}_\text{L}\cdot \vec{r}_j} \hat{P}^j_{13} +\hat{F}^j_{12},
\label{eq:P_12}
}
where~$\dot{}$ the time derivative
and~$\bar{P}^j_1\approx1$,
$\gamma_{pq}$ are decay constants for atomic transitions $\ket{p}\leftrightarrow\ket{q}$,   $N_1$ and $N_2$ are numbers of atoms in memory and dispersion compensation subensembles. 

In Eq.~(\ref{eq:a_c}),
we include coupling to the $\ket5\leftrightarrow\ket6$
transition for dispersion compensation.
Our proposal for dispersion interaction compensates a round-trip phase shift for different spectral components of the input field,
thus keeping them under impedance-matching conditions.
We first describe mathematically the interaction between the lasers and all atomic levels,
then linearize these equations
and finally describe,
based on these linearized equations,
how a subensemble of $\Lambda$-type atoms serves this purpose.

The equations for driven transitions between levels~$\ket4$, $\ket5$ and $\ket6$ are
\begin{align}
\dot{\hat{P}}^j_{64}
=& -(\gamma_{46}-\text{i}\Delta_{\text{L}'})\hat{P}^j_{64}+\text{i}\Omega_{\text{L}'}\text{e}^{-\text{i}\vec{k}_{\text{L}'}\cdot \vec{r}_j} \bar{P}^j_4
+\hat{F}^j_{64}, \label{eq:r46}\\
\dot{\hat{P}}^j_{56}
=& -(\gamma_{56}-\text{i}(\Delta_{\text{L}'}+\delta ) )\hat{P}^j_{56}-i \hat{P}^j_{54}\Omega_{\text{L}'}\text{e}^{\text{i}\vec{k}_{\text{L}'}\cdot \vec{r}_j} \nonumber \\ &+ i g_{56}^{j*}\text{e}^{\text{i}\vec{k}_\text{c}\cdot \vec{r}_j} \hat{a}_\text{c} \bar{P}^j_{6}
+\hat{F}^j_{56}, \label{eq:r56}\\
\dot{\hat{P}}^j_{54}
=& -(\gamma_{45}+\text{i}\delta )\hat{P}^j_{54}-i \hat{P}^j_{56}\Omega_{\text{L}'}\text{e}^{-\text{i}\vec{k}_{\text{L}'}\cdot \vec{r}_j} 
\nonumber\\
&+\text{i}g_{56}^{j*}\text{e}^{\text{i}\vec{k}_\text{c}\cdot \vec{r}_j}\hat{a}_\text{c} \hat{P}_{64}^j
+\hat{F}^j_{54}, 
\label{eq:r45}
\end{align}
where this dispersion-compensation ensemble has near-unity population in level~$\ket4$ so $\bar{P}^j_4\approx1$ while 
$\bar{P}^j_{6}(t)\cong\frac{\Omega_{\text{L}'}^2(t)}{\gamma_{56}^2+\Delta_{\text{L}'}^2}\bar{P}^j_{4}$. 
Additionally, $\Omega_\text{L}$ and $\Omega_{\text L'}$ are kept constant during write-in stage in Eqs.~\eqref{eq:P_13}--\eqref{eq:r45}.


For~$\Delta_{\text{L}'}\gg \max\{\gamma_{46},\gamma_{56},\Omega_{\text{L}'}\}$,
the noise term
$\hat{F}^j_{64}(t)\propto\gamma_{46}$
is negligible
so we adiabatically eliminate the left-hand side of Eq.~(\ref{eq:r46});
i.e., $\hat{P}^j_{64}$ is effectively constant.
Thus, we replace~$\hat{P}^j_{64}$
by $\bar{P}^j_{64}$ in ~(\ref{eq:r45}),
so Eqs.~(\ref{eq:a_c}--\ref{eq:P_12})
combined with Eqs.~(\ref{eq:r46}--\ref{eq:r45}),
are collectively linearized and thus amenable to algebraic analysis via a Fourier transform with respective component $\omega$.
We solve Eqs.~(\ref{eq:acaoutain}--\ref{eq:r45}),
bearing in mind that the input field is mapped onto inhomogeneously broadened  spin coherence of the transition $\ket1\leftrightarrow\ket2$~\cite{SMoiseev2013},
to obtain $\hat{{a}}_\text{out}(\omega)
=r(\omega)\hat{{a}}_\text{in}(\omega)$
for reflectivity $r(\omega)=\xi_+(\omega)/\xi_-(\omega)$ for
\fla{
\xi_\pm(\omega)
:=\kappa/2 \mp \color{black}\beta_1(\omega) \pm\color{black}\text{i}\beta_2(\omega)\pm\text{i}\omega,
\label{xi}
}
with 

\begin{equation}
\beta_{1}(\omega)=i \sum^{N_1}_{j=1}\abs{g_{13}^j}^2 \frac{ \bar{P}^j_{1}} {(\Delta_\text{L}-\omega-\text{i}\gamma_{13})}+\beta_{1,\text{ab}}(\omega),
\label{eq:beta1}
\end{equation}

\noindent
characterizing spectral response of the memory, where the term  
\fla{
\beta_{1,\text{ab}}(\omega)
&=
\sum^{N_1}_{j=1}  \frac{ \text{i}\abs{g_{13}^j}^2 \bar{P}^j_{1}\frac{\abs{\Omega_\text{L}}^2}{(\Delta_L-\text{i}\gamma_{13})^2}}{(\delta^j_{21}-\omega-\text{i}\gamma_{12})-\frac{\abs{\Omega_{L}}^2 }{(\Delta_L-\text{i}\gamma_{13})}}
\label{beta:1ab}
}
is responsible for the coherent Raman absorption of the signal field. Function
\fla{
\label{eq:beta2}
\beta_2(\omega)=
\bar{P}_4
\sum^{N_2}_{j=1}
\frac{|g_{56}^j|^2\Omega_{\text{L}'}^2\left(\frac{1}{\Delta_{\text{L}'}+i \gamma_{46}}-\frac{\omega-\delta+\text{i}\gamma_{45}}{\Delta_{\text{L}'}^2+\gamma_{46}^2}\right)}
{\Omega_{\text{L}'}^2+(\Delta_{\text{L}'}+\delta -\omega-\text{i}\gamma_{56})(\omega-\delta+\text{i} \gamma_{45})}
}

\noindent
characterizes dispersion-compensation subensembles.

To achieve our goal of broadband quantum memory via zero spectral dispersion,
we employ impedance matching \cite{AfzeliusPRA2010,Moiseev2010b,SMoiseev2013}.
Impedance matching requires that the reflected beam has zero intensity so reflection~$\abs{r(\omega)}^2$ needs to be close to zero.
Whereas both $\beta_1(\omega)$ and $\beta_2(\omega)$ are complex, our intuition is based on approximately real $\beta_1(\omega)$ and $\beta_2(\omega)$ in narrow frequency domain around $\omega \approx 0$.   
The function $\beta_1(\omega)$ is almost real in the neighbourhood of $\omega=0$ bounded by $|\omega|\ll\Delta_\text{in}$.
Similarly, $\beta_2(\omega)$ is almost real if $\abs{\omega}\ll\delta$.
Therefore, if we restrict frequencies to be in the domain
$\abs{\omega}\ll\text{min}(\delta,\Delta_\text{in})$,
the memory response is wholly absorptive; and absorption by dispersive media can be neglected. 
The zero-reflectivity condition is satisfied if
$\beta_1(\omega) \approx \kappa/2$, 
and dispersion compensation satisfies $\beta_2(\omega)\approx-\omega$.
In a bigger frequency range we treat both functions as complex and demonstrate numerically how bandwidth of the memory is extended by tuning control parameters of $\beta_2(\omega)$.


After the signal pulse is absorbed and the control fields are adiabatically switched off as discussed in \cite{EMoiseev2013}, spectral components of the incoming field are stored on the  long-lived coherence $\hat P^j_{12}$ of the corresponding resonant atoms (corrected by AC-Stark shift):
\begin{align}
\hat P^j_{12}(t)&=\frac{i\sqrt{2\pi \kappa}\Omega^{*}_{\text{L}}g_{13}^{j*}
\hat{{a}}_\text{in}(\delta^j_{21}) 
e^{-i(\delta^j_{21}-i\gamma_{12})t}
}
{\xi_-(\delta^j_{21})\left(\Delta_{\text{L}}-\delta^j_{21}-i{\gamma}_{13} \right)}.
\label{P12:st}
\end{align}
Here Rabi frequencies $\Omega_{\text{L}}$ and $\Omega_{\text{L'}}$ are constant with the values taken from the write-in stage.
The presence of inhomogeneous broadening with characteristic width $\Delta_{\text{in}}$ will reset the collective coherence after time $\sim 1/\Delta_{\text{in}}$ allowing temporal multiplexing.

\section{Retrieval}
To retrieve the light pulse from the memory, collective coherence must be restored. 
Restoring can be done by various methods, such as inversion of the inhomogeneous broadening $\delta^{j}_{21} \rightarrow -\delta^{j}_{21}$ as in the controllable reversal of inhomogeneous broadening (CRIB) protocol \cite{SMoiseev2013}, applying a pair of $\pi$ pulses on $\ket{1}\leftrightarrow\ket{2}$ separated by time interval $\tau$ as in the ROSE protocol \cite{Bonarota2014} or by exploiting the presence of a comb structure in inhomogeneous broadening profile as in the atomic frequency comb (AFC) protocol  \cite{Riedmatten2008}.

Once the rephasing of the coherence is realized by either CRIB, ROSE or AFC at  time $t=\tau$, the control fields $\Omega_\text{L}$, $\Omega_{\text{L}'}$  are switched on.
Switching of the inhomogeneous broadening in CRIB protocol results $\text{i}\delta_{21}$ \eqref{eq:P_12} becoming negative, whereas use of AFC or ROSE will retain the positive sign. 
Subsequent dynamics over time $t>\tau$ is described by
\fla{
\dot{\hat{a}}_\text{c} =& -\frac{\kappa}{2} \hat{a}_\text{c}- \text{i}\sum^{N_1}_{j=1} \left( g_{13}^j \hat{P}^j_{13}+g_{56}^j \hat{P}^j_{56}  \right)\text{e}^{-\text{i}\vec{k}_\text{c} \cdot\vec{r}_j}
\nonumber \\
&+\sqrt{\kappa}\hat{b}_\text{in},
\label{eqn:a_c}\\
\dot{\hat{P}}^j_{12} =& -(\gamma_{12}\mp\text{i}\delta_{21}^j ) \hat{P}^j_{12} - i\Omega_\text{L}^*\text{e}^{-\text{i}\vec{k}_\text{L}\cdot \vec{r}_j} \hat{P}^j_{13} \nonumber \\
&+\hat{F}^j_{12}
+ \delta(t-\tau)\hat{P}^j_{12}(\tau), 
\label{Eq:P12:time}
}
where $\hat{b}_\text{in}$ is the cavity input field operator at the echo emission stage. Further on the input field assumed to be in a vacuum state.  
Together with Eqs.~(\ref{eq:P_13}), (\ref{eq:r46})--(\ref{eq:r45}), they form a system which is the same as Eqs.~\eqref{eq:a_c}--\eqref{eq:r45} with different initial conditions.
The additional term in Eq.~(\ref{Eq:P12:time}), namely $\delta(t-\tau)\hat{P}^j_{12}(\tau)$, describes the initial atomic coherence  at $t=\tau$ given by Eq.~(\ref{P12:st}).

Similarly, as before,
we solve this system of equations for times $t>\tau$ \cite{SMoiseev2013} and get a beam-splitter type relation for the output field in the limit $\gamma_{12}\tau\ll 1$: 
\fla{
&\hat{a}_{\text{out }\mp}(t)=\frac{e^{-2\gamma_{12}\tau}}{\sqrt{2\pi}}\int d\omega e^{-i\omega(t-2\tau)} T_{\mp}(\omega) \hat{a}_\text{in}(\mp\omega) \nonumber \\ &+ \int d \omega \frac{e^{-i\omega t}}{\sqrt{2\pi}}\sqrt{1-e^{-4\gamma_{12}\tau}\abs{T_{\mp}(\omega)}^2} \frac{r(\omega)}{|{r(\omega)}|}\hat{{b}}_\text{in}(\omega),
\label{a_out}
}
where $T_{\mp}(\omega)$ is a memory response function for CRIB (AFC or ROSE)  and is related to efficiency for a given spectral spectral component as $\wp(\omega) = \abs{T_\mp(\omega)}^2$.

 Below we perform noise analysis allowing us to assume that the bath field is in vacuum and that the memory has low noise. 
If the rephasing is time-reversed as in CRIB the spectral storage function is
\fla{
T_{-}(\omega)=-\frac{2\kappa \Re e[\beta_{1,\text{ab}}(\omega)]}{
\xi_{-}(\omega)\xi_{-}(-\omega)}.
\label{T_crib}
}
This function governs emission of an echo signal at $t=2\tau$ with an inverted spectrum and time-reversed shape \cite{EMoiseev2013}. 
Specifically, if inhomogeneous broadening is symmetric, we get $\xi_{-}(-\omega)=\xi_{-}^*(\omega)$ and  $\xi_{-}(\omega)\xi_{-}(-\omega)=\abs{\xi_{-}(\omega)}^2$, which  leads to time reversibility of this protocol without parasitic phase distortion.

Similar calculations performed for AFC and ROSE protocols \cite{Riedmatten2008,Bonarota2014,Simon2010a} lead to the spectral storage function 
\fla{
T_{+}(\omega)=-\frac{2\kappa \Re e[\beta_{1,\text{ab}}(\omega)]}{
\xi_{-}^2(\omega)}.
\label{T_out2}
}
This storage function does not fully preserve time reversibility \cite{Moiseev2012} and could lead to unwanted additional spectral modulation of the emitted echo $\hat{a}_\text{out}(\omega)\sim \hat{a}_\text{in}(\omega)e^{-2i\phi(\omega)}$, where $\tan\phi(\omega)=\Im m[\xi_{-}(\omega)]/\Re e[\xi_{-}(\omega)]$. Thus, the AFC and ROSE protocols produce the signal field with the same spectral efficiency $\wp(\omega)$, but they are more sensitive to the presence of a spectral dispersion.
Knowing the output field from the resonator after retrieval Eq.~\eqref{a_out} we can estimate efficiency of our memory taking into account the CRIB rephasing procedure for the sake of simplicity.

\noindent

\begin{figure}
  \includegraphics[width=0.49\columnwidth]{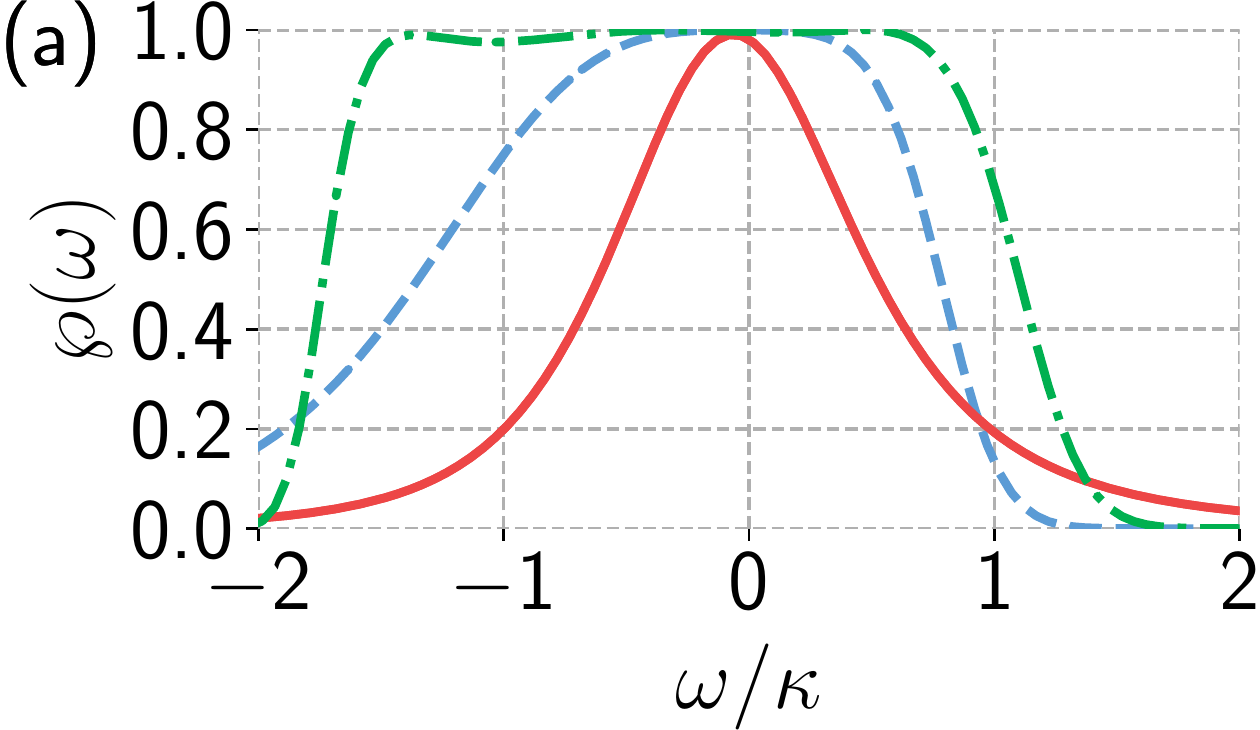}
  \includegraphics[width=0.49\columnwidth]{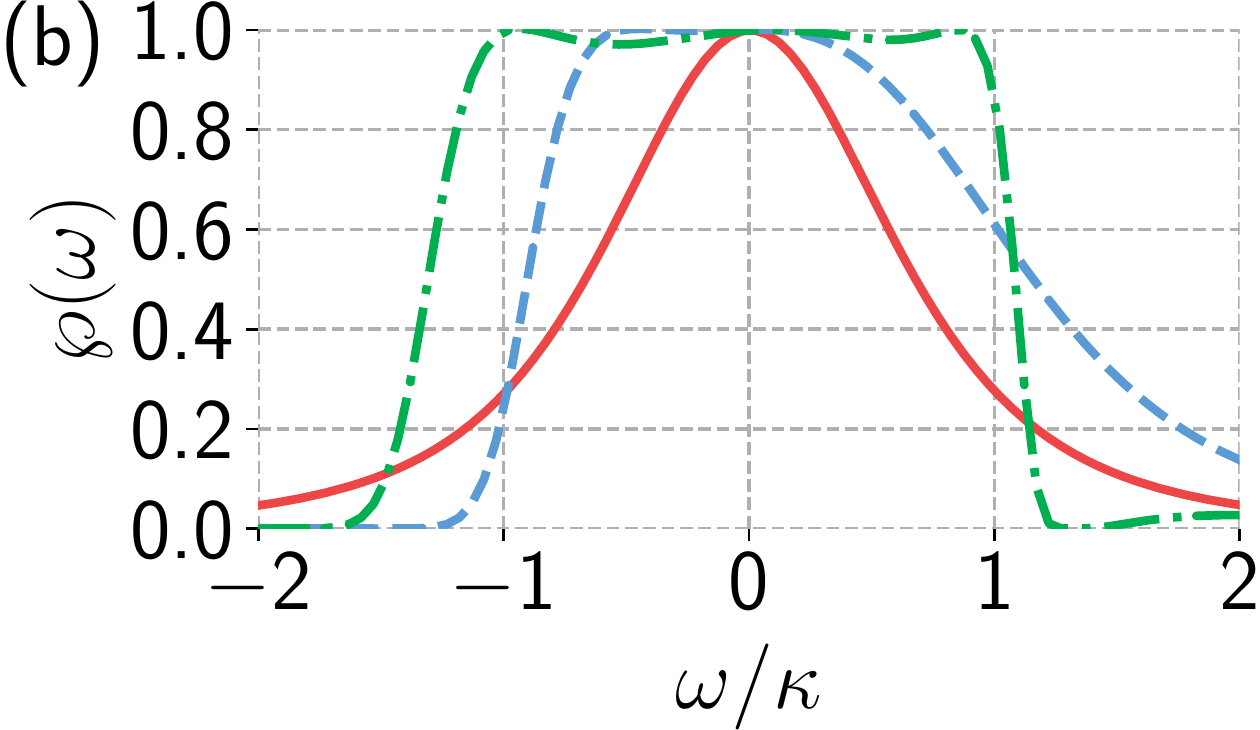}
\caption[Numerical simulation]{%
Spectral efficiency~$\wp(\omega)$ vs scaled frequency~$\omega/\kappa$
for off-resonant Raman impedance-matched quantum memory (\textcolor{red1}{\textbf{red}} solid) without dispersion compensation $\Omega_{\text{L}'}=0$ and without phase modulation $n_\text{c}=0$;
with dispersion compensation and without phase modulation $n_\text{c}=0$ (\textcolor{blue1}{\textbf{blue}} dashed);
with phase-modulated dispersion compensation  (\textcolor{green1}{\textbf{green}} dot-dashed).
Two subplots are calculated for different inhomogeneous broadening shapes (a)~Lorentzian and (b)~uniform rectangular.
The chosen parameters,
normalized to $\kappa=1$,
are
$\gamma_{12}=\gamma_{45}=10^{-4}$,
$\gamma_{13}=\gamma_{46}=\gamma_{56}=0.1$. 
We also choose
(a)~$\Omega_\text{L}=9.65$,
$\Omega_{\text{L}'}=5.5$,
$\Delta_\text{L}=100$,
$\Delta_{\text{L}'}=100$,
$\delta=-3.55$, $N_1 \bar{P}_1\abs{g_{13}}^2=750$, $N_2 \bar{P}_4\abs{g_{56}}^2=1150$, $\Delta_{\text{SB}}=\pm 4.6$, $\alpha=0.53$  and
$\Delta_\text{in}=10$.
(b)~$\Omega_\text{L}=10.7$,
$\Omega_{\text{L}'}=5.75$,
$\Delta_\text{L}=150$,
$\Delta_{\text{L}'}=100$,
$\delta=-5.6$, $N_1 \bar{P}_1 \abs{g_{13}}^2=480$, $N_2 \bar{P}_4 \abs{g_{56}}^2=480$, $\Delta_{\text{SB}}=\pm 3.0$, $\alpha=0.53$  and
$\Delta_\text{in}=14$.
}
\label{fig:reflection}
\end{figure}
\section{Memory efficiency}
In Fig.~\ref{fig:reflection},
we present numerical simulations of spectral efficiency $\wp(\omega)$  for two shapes of inhomogeneous broadening, namely Lorentzian and rectangular uniform. The latter is simpler to achieve in experiment by application of a linear magnetic field gradient, whereas the Lorentzian is easier for theoretical analysis. 
See Appendix \ref{app:IB} for more details on inhomogeneous broadening shapes.

Spectral dispersion is compensated either by a single quasi-CW control field or by phase-modulated control field. 
The goal of phase modulation is to create side bands of control field, 
which operate as additional independent dispersion-compensation fields. Modified $\beta_2(\omega)$ under an action of phase-modulated control field are described in Appendix \ref{app:pm}. 

As seen in Fig.~\ref{fig:reflection},
the introduced dispersion compensation $\beta_2(\omega)$ significantly extends the quantum memory bandwidth.
In particular, for a Lorentzian line shape at a given minimum spectral efficiency threshold $\wp=0.9$, we observe a bandwidth expansion
3.32 times for a single control field and 5.85 times for a phase-modulated control field
in comparison with no dispersion compensation~\cite{Moiseev2010b}.
For higher minimum spectral efficiency threshold $\wp=0.95$, we get bandwidth expansions 4.52 times and  9.26 times, respectively.
In both these cases, increased bandwidth of quantum memory exceeds the linewidth of an empty resonator $\delta\omega_{\text{qm}}>\kappa$.

Similar memory bandwidth extension is possible to reach for uniform inhomogeneous broadening. 
However, there is one important difference leading to it's advantage in terms of experimental feasibility over Lorentz.  
If we apply an external magnetic field to broaden the transition $\ket{1}\leftrightarrow\ket{2}$, we inevitably cause inhomogeneous broadening on the transition $\ket{4}\leftrightarrow\ket{5}$. We demonstrate that our dispersion compensation  remains feasible for uniform inhomogeneous broadening as elaborated in Appendix \ref{Appx:InhomoDispComp}.  In contrast, dispersion-compensation media with huge Lorentzian or Gaussian inhomogeneous broadening requires impractically high collective coupling constant to achieve sharp enough compensating dispersion. 
Thus, we have shown that CRIB is feasible for uniform rectangular inhomogeneous broadening, whereas ROSE and AFC with non-uniform broadening are impractical.
  
\section{Noise}
Now we incorporate noise in our model to establish that our scheme works well even in the presence of noise.
We have justified neglecting noise associated with transition $\ket4\leftrightarrow\ket6$ so here we discuss other noises that can be neglected.
Then we study nonnegligible noise and show that our scheme is robust despite this noise.

First we discuss which noise terms can be neglected.
As our scheme involves two subensembles,
exemplified by additivity of interaction between the cavity and the dispersion-compensation medium~(\ref{eq:a_c}),
we separate consideration of noise processes on the $\ket5\leftrightarrow\ket4$ and~$\ket5\leftrightarrow\ket6$
transitions from consideration of noise processes on the $\ket1\leftrightarrow\ket2$ and~$\ket1\leftrightarrow\ket3$.
Noise for $\ket1\leftrightarrow\ket2$ and~$\ket1\leftrightarrow\ket3$
are small as the coherence time for
the $\ket1\rightarrow\ket2$ transition is large and the initial population in level $\ket{1}$ is large~\cite{FleischhauerPRA2013, Nunn2017, Hsiao2018}.
Moreover, quantum noise induced by the four-wave mixing \cite{PolzikCirac2011,  Moiseev2020} in a $\Lambda$ scheme are suppressed by presence of a high-Q cavity ($\kappa\ll\omega_{21}$).

Time-reversal character of Eqs.~(\ref{eq:a_c})--(\ref{eq:P_12}) and (\ref{eqn:a_c})--(\ref{Eq:P12:time}) together with the spontaneous nature of the noise \eqref{eq:r46}--\eqref{eq:r45} leads to the same noise being generated at the write-in and read-out stages if control parameters are maintained the same.
Therefore, by considering only a single (write-in) stage, we determine half of the total noise contribution.

We now solve Eqs.~(\ref{eq:a_c}) and~(\ref{eq:r46}--\ref{eq:r45})
with the remaining quantum Langevin noise terms~$\hat{F}^j_{54}$ and~$\hat{F}^j_{56}$ intact
to determine the effect of noise on the dispersion-compensating medium. 
The output vacuum-noise spectrum is $\langle n\rangle =\color{black}\mean{\hat{a}_{\text{out}}^{\dagger}(t')\hat{a}_{\text{out}}(t)}=\int\text{d}\omega S(\omega)\text{e}^{-\text{i}\omega(t-t')}$ for
\fla{
S(\omega)=&
\left|\frac{\kappa g_{56}}{\xi_-(\omega)}\right|^2 \sum^{N_2}_{j=1}\Bigg(\frac{\gamma_{56} \bar{P}^j_6}{\abs{\Xi(\omega)}^2}
    \nonumber\\
&+\frac{  \abs{\Omega_{\text{L}'}\color{black}}^2 (\gamma_{54} \bar{P}^j_{4} +(\gamma_{46}+\gamma_{56})\bar{P}^j_6  ) }{\abs{\Gamma(\omega) \Xi(\omega)}^2} \Bigg)\label{nspect},
}
with
$\Gamma(\omega):=\gamma_{45}-\text{i}\delta +\text{i}\omega$
and
$\Xi(\omega)
:=\gamma_{56}+\text{i}\Delta_{\text{L}'} -\text{i}\omega+\abs{\Omega_{\text{L}'}\color{black}}^2/\Gamma(\omega)$.

\begin{figure}
  \centering
  \includegraphics[width=1\columnwidth]{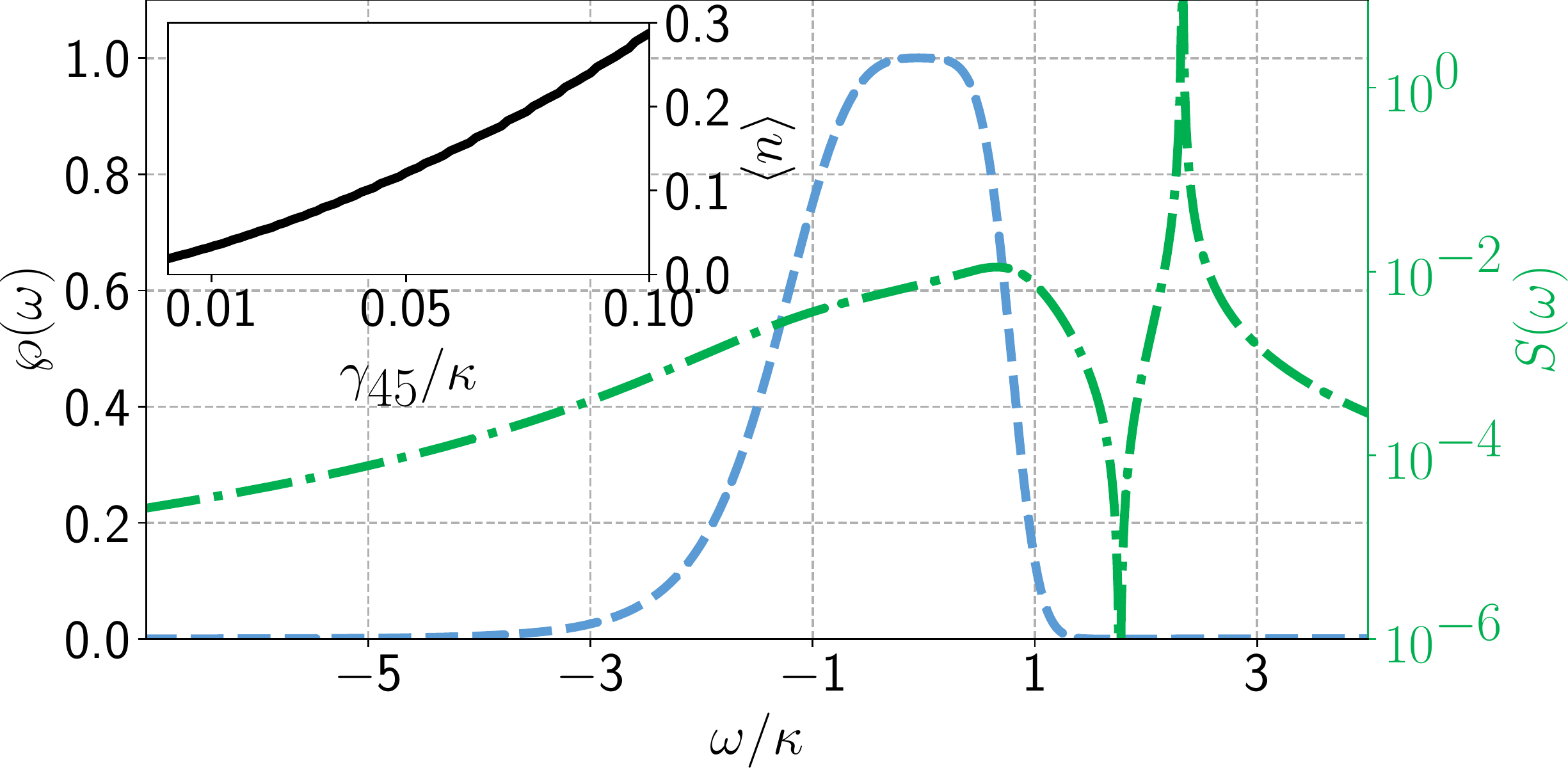}
  \caption{ Noise spectrum for spontaneously emitted photons  (\textcolor{green1}{\textbf{green}} dot-dashed) with broadened (\textcolor{blue1}{\textbf{blue}} dashed curve) spectral efficiency  profiles.  
Inset: Number of noise photons $\langle {n} \rangle$ as a function of $\gamma_{45}$. 
The chosen parameters are the same as in Fig.~\ref{fig:reflection}(a), except that we assumed nonideal depopulation of level $\ket{6}$, which has 5\% of population.
\label{fig:WhiteCavity:Noises:Raman} }
\end{figure}

To determine the noise performance caused by additional dispersion interaction,  we consider experimentally feasible medium explained in details below.
The noise spectrum and
spectral efficiency in Fig.~\ref{fig:WhiteCavity:Noises:Raman} 
\newcommand{\shiftleft}[2]{\makebox[0pt][r]{\makebox[#1][l]{#2}}}
show that our impedance-matched quantum memory has an impressively low noise level of $\sim 10^{-2}$ photons across the bandwidth with 90\% efficiency despite this noise,
and this performance enables operation at the single-photon level with high fidelity~\cite{Lobino2009}.

\section{Discussions and conclusions}
Success for our scheme requires a highly coherent $\ket4\leftrightarrow\ket5$ transition so $\delta/\gamma_{45}\gg1$,
and noise-power dependence on~$\gamma_{45}$
is presented in the inset of Fig.~\ref{fig:WhiteCavity:Noises:Raman}. 
Thus, our procedure requires a three-level atom with a sufficiently long coherence time between the two lowest atomic levels and inhomogeneous broadening of transitions between these lowest levels with a third level such that subensembles can be independently manipulated by laser interaction. 

These stringent conditions,
including coherence between the two lowest levels,
are met by the $\text{Er}^{3+}$ ion with the telecom-compatible
$^4\text{I}_{15/2}\leftrightarrow ^4\text{I}_{13/2}$ transition 
and is thus a natural choice for optical quantum memory~\cite{Craiciu2019, Gisin2011,Tittel2019PRAppl}.
At large magnetic field
effective hole burning, long coherence time and ground-state spin polarization of $^{167}\text{Er}^{3+}$ are achieved~\cite{2017-Rancic-NP}.
Thus, this crystal is promising for realizing the broadband impedance-matched telecom quantum memory.

Now that we have recommended an appropriate material,
we proceed to explore possible cavity designs that could provide requisite impedance matching. 
The most straightforward solution would be to place a rare-earth-doped crystal into a Fabry-Perot resonator~\cite{Jobez2014, 2017-Minnegaliev-QE}. 
Although inherited large intra-cavity losses even for anti-reflective coated crystals, makes them unfavourable for our task.

Two more designs are the monolithic Fabry-Perot resonator,
where crystal surfaces are reflection-coated to create a Fabry-Perot resonator within the crystal~\cite{Goto2010, SabooniPRL2013},
and the whispering-gallery mode resonator~\cite{McAuslan2011PRA},
where the whispering-gallery mode is milled from the rare-earth-doped crystal.  
The losses for both of these designs are bounded by  bulk $\text{Y}_2\text{SiO}_5$ losses per round trip and estimated to be $7\cdot 10^{-4} \text{cm}^{-1}$ for visible light~\cite{Goto2010}.
These losses are due to material absorption~\cite{Goto2010}
and dominate over fundamental losses due to Rayleigh and Brillouin scattering~\cite{Grudinin2007}. 
The resulting quality factors of monolithic and whispering-gallery-mode resonators are comparable.
Experimental simplicity of monolithic design makes this approach preferable.

Finally, we assess feasibility for experimental realization with a 9~mm monolithic cavity created from $\text{Y}_2\text{SiO}_5$ crystal doped with isotopically pure $^{167}\text{Er}^{+3}$~\cite{Goto2010}.
We propose burning the sample into a memory ensemble with optical linewidth 10~MHz and a dispersion-compensation ensemble with linewidth 1~MHz~\cite{2017-Rancic-NP}.
According to our model, the dispersion-compensation subensemble should contain as much as $55$\% of the total number of atoms in the memory subensemble.
Our design can provide an inverse photon lifetime of $\kappa/2\pi \sim 6.3$  MHz with cavity volume 1~mm$^3$  and an average single-photon cavity-atom coupling constant of $g/2\pi\sim$100~Hz together with an ion density of $2\times10^{12}$~mm$^{-3}$. 

For this choice of materials and parameters,
our scheme promises a 90\% bandwidth extension from the previous prediction of 4~MHz in an impedance-matched scheme~\cite{SMoiseev2013}
to 10~MHz in our case  with additional noise limited to being less than $0.01$ photons per pulse.
Taking into account the one second coherence lifetime of the hyperfine levels \cite{2017-Rancic-NP} the resultant delay-bandwidth product could reach $\sim 10^{7}$ in telecom-compatible material.
The use of a phase-modulated control field may open the avenue for extending the bandwidth up to 40~MHz.

In this work we propose an optical quantum memory protocol with zero spectral dispersion engineering within a high-Q cavity. 
Our proposal is based on extending of the impedance matching to a broader frequency range by compensating for dispersion. 
We show that dispersion compensation increases 
the quantum storage bandwidth by more than an order of magnitude. 
By adopting our approach,
a large delay-bandwidth product and a long storage time in solid-state media should be feasible.

\acknowledgments
SAM thanks support within framework project \# 00075-02-2020-051/1 from 02.03.2020.
BCS appreciates financial support from NSERC.
\bibliography{lib}

\begin{thebibliography}{49}%
\makeatletter
\providecommand \@ifxundefined [1]{%
 \@ifx{#1\undefined}
}%
\providecommand \@ifnum [1]{%
 \ifnum #1\expandafter \@firstoftwo
 \else \expandafter \@secondoftwo
 \fi
}%
\providecommand \@ifx [1]{%
 \ifx #1\expandafter \@firstoftwo
 \else \expandafter \@secondoftwo
 \fi
}%
\providecommand \natexlab [1]{#1}%
\providecommand \enquote  [1]{``#1''}%
\providecommand \bibnamefont  [1]{#1}%
\providecommand \bibfnamefont [1]{#1}%
\providecommand \citenamefont [1]{#1}%
\providecommand \href@noop [0]{\@secondoftwo}%
\providecommand \href [0]{\begingroup \@sanitize@url \@href}%
\providecommand \@href[1]{\@@startlink{#1}\@@href}%
\providecommand \@@href[1]{\endgroup#1\@@endlink}%
\providecommand \@sanitize@url [0]{\catcode `\\12\catcode `\$12\catcode
  `\&12\catcode `\#12\catcode `\^12\catcode `\_12\catcode `\%12\relax}%
\providecommand \@@startlink[1]{}%
\providecommand \@@endlink[0]{}%
\providecommand \url  [0]{\begingroup\@sanitize@url \@url }%
\providecommand \@url [1]{\endgroup\@href {#1}{\urlprefix }}%
\providecommand \urlprefix  [0]{URL }%
\providecommand \Eprint [0]{\href }%
\providecommand \doibase [0]{http://dx.doi.org/}%
\providecommand \selectlanguage [0]{\@gobble}%
\providecommand \bibinfo  [0]{\@secondoftwo}%
\providecommand \bibfield  [0]{\@secondoftwo}%
\providecommand \translation [1]{[#1]}%
\providecommand \BibitemOpen [0]{}%
\providecommand \bibitemStop [0]{}%
\providecommand \bibitemNoStop [0]{.\EOS\space}%
\providecommand \EOS [0]{\spacefactor3000\relax}%
\providecommand \BibitemShut  [1]{\csname bibitem#1\endcsname}%
\let\auto@bib@innerbib\@empty
\bibitem [{\citenamefont {Lvovsky}\ \emph {et~al.}(2009)\citenamefont
  {Lvovsky}, \citenamefont {Sanders},\ and\ \citenamefont {Tittel}}]{QMReview}%
  \BibitemOpen
  \bibfield  {author} {\bibinfo {author} {\bibfnamefont {A.~I.}\ \bibnamefont
  {Lvovsky}}, \bibinfo {author} {\bibfnamefont {B.~C.}\ \bibnamefont
  {Sanders}}, \ and\ \bibinfo {author} {\bibfnamefont {W.}~\bibnamefont
  {Tittel}},\ }\href@noop {} {\bibfield  {journal} {\bibinfo  {journal} {Nat.
  photonics}\ }\textbf {\bibinfo {volume} {3}},\ \bibinfo {pages} {706}
  (\bibinfo {year} {2009})}\BibitemShut {NoStop}%
\bibitem [{\citenamefont {Hammerer}\ \emph {et~al.}(2010)\citenamefont
  {Hammerer}, \citenamefont {S\o{}rensen},\ and\ \citenamefont
  {Polzik}}]{Hammerer2010}%
  \BibitemOpen
  \bibfield  {author} {\bibinfo {author} {\bibfnamefont {K.}~\bibnamefont
  {Hammerer}}, \bibinfo {author} {\bibfnamefont {A.~S.}\ \bibnamefont
  {S\o{}rensen}}, \ and\ \bibinfo {author} {\bibfnamefont {E.~S.}\ \bibnamefont
  {Polzik}},\ }\href {\doibase 10.1103/RevModPhys.82.1041} {\bibfield
  {journal} {\bibinfo  {journal} {Rev. Mod. Phys.}\ }\textbf {\bibinfo {volume}
  {82}},\ \bibinfo {pages} {1041} (\bibinfo {year} {2010})}\BibitemShut
  {NoStop}%
\bibitem [{\citenamefont {Heshami}\ \emph {et~al.}(2016)\citenamefont
  {Heshami}, \citenamefont {England}, \citenamefont {Humphreys}, \citenamefont
  {Bustard}, \citenamefont {Acosta}, \citenamefont {Nunn},\ and\ \citenamefont
  {Sussman}}]{Khabat2016}%
  \BibitemOpen
  \bibfield  {author} {\bibinfo {author} {\bibfnamefont {K.}~\bibnamefont
  {Heshami}}, \bibinfo {author} {\bibfnamefont {D.~G.}\ \bibnamefont
  {England}}, \bibinfo {author} {\bibfnamefont {P.~C.}\ \bibnamefont
  {Humphreys}}, \bibinfo {author} {\bibfnamefont {P.~J.}\ \bibnamefont
  {Bustard}}, \bibinfo {author} {\bibfnamefont {V.~M.}\ \bibnamefont {Acosta}},
  \bibinfo {author} {\bibfnamefont {J.}~\bibnamefont {Nunn}}, \ and\ \bibinfo
  {author} {\bibfnamefont {B.~J.}\ \bibnamefont {Sussman}},\ }\href {\doibase
  10.1080/09500340.2016.1148212} {\bibfield  {journal} {\bibinfo  {journal} {J.
  Mod. Opt.}\ }\textbf {\bibinfo {volume} {63}},\ \bibinfo {pages} {2005}
  (\bibinfo {year} {2016})}\BibitemShut {NoStop}%
\bibitem [{\citenamefont {Chanelière}\ \emph {et~al.}(2018)\citenamefont
  {Chanelière}, \citenamefont {Hétet},\ and\ \citenamefont
  {Sangouard}}]{CHANELIERE2018}%
  \BibitemOpen
  \bibfield  {author} {\bibinfo {author} {\bibfnamefont {T.}~\bibnamefont
  {Chanelière}}, \bibinfo {author} {\bibfnamefont {G.}~\bibnamefont {Hétet}},
  \ and\ \bibinfo {author} {\bibfnamefont {N.}~\bibnamefont {Sangouard}},\ }in\
  \href {\doibase https://doi.org/10.1016/bs.aamop.2018.02.002} {\emph
  {\bibinfo {booktitle} {Advances In Atomic, Molecular, and Optical
  Physics}}},\ Vol.~\bibinfo {volume} {67},\ \bibinfo {editor} {edited by\
  \bibinfo {editor} {\bibfnamefont {E.}~\bibnamefont {Arimondo}}, \bibinfo
  {editor} {\bibfnamefont {L.~F.}\ \bibnamefont {DiMauro}}, \ and\ \bibinfo
  {editor} {\bibfnamefont {S.~F.}\ \bibnamefont {Yelin}}}\ (\bibinfo
  {publisher} {Academic Press},\ \bibinfo {year} {2018})\ Chap.~\bibinfo
  {chapter} {2}, pp.\ \bibinfo {pages} {77--150}\BibitemShut {NoStop}%
\bibitem [{\citenamefont {Razavi}\ \emph {et~al.}(2009)\citenamefont {Razavi},
  \citenamefont {Piani},\ and\ \citenamefont {L{\"{u}}tkenhaus}}]{Razavi2009}%
  \BibitemOpen
  \bibfield  {author} {\bibinfo {author} {\bibfnamefont {M.}~\bibnamefont
  {Razavi}}, \bibinfo {author} {\bibfnamefont {M.}~\bibnamefont {Piani}}, \
  and\ \bibinfo {author} {\bibfnamefont {N.}~\bibnamefont {L{\"{u}}tkenhaus}},\
  }\href {\doibase 10.1103/PhysRevA.80.032301} {\bibfield  {journal} {\bibinfo
  {journal} {Phys. Rev. A}\ }\textbf {\bibinfo {volume} {80}},\ \bibinfo
  {pages} {032301} (\bibinfo {year} {2009})}\BibitemShut {NoStop}%
\bibitem [{\citenamefont {Simon}\ \emph {et~al.}(2007)\citenamefont {Simon},
  \citenamefont {de~Riedmatten}, \citenamefont {Afzelius}, \citenamefont
  {Sangouard}, \citenamefont {Zbinden},\ and\ \citenamefont
  {Gisin}}]{Simon2007}%
  \BibitemOpen
  \bibfield  {author} {\bibinfo {author} {\bibfnamefont {C.}~\bibnamefont
  {Simon}}, \bibinfo {author} {\bibfnamefont {H.}~\bibnamefont
  {de~Riedmatten}}, \bibinfo {author} {\bibfnamefont {M.}~\bibnamefont
  {Afzelius}}, \bibinfo {author} {\bibfnamefont {N.}~\bibnamefont {Sangouard}},
  \bibinfo {author} {\bibfnamefont {H.}~\bibnamefont {Zbinden}}, \ and\
  \bibinfo {author} {\bibfnamefont {N.}~\bibnamefont {Gisin}},\ }\href
  {\doibase 10.1103/PhysRevLett.98.190503} {\bibfield  {journal} {\bibinfo
  {journal} {Phys. Rev. Lett.}\ }\textbf {\bibinfo {volume} {98}},\ \bibinfo
  {pages} {190503} (\bibinfo {year} {2007})}\BibitemShut {NoStop}%
\bibitem [{\citenamefont {Afzelius}\ and\ \citenamefont
  {Simon}(2010)}]{AfzeliusPRA2010}%
  \BibitemOpen
  \bibfield  {author} {\bibinfo {author} {\bibfnamefont {M.}~\bibnamefont
  {Afzelius}}\ and\ \bibinfo {author} {\bibfnamefont {C.}~\bibnamefont
  {Simon}},\ }\href {\doibase 10.1103/PhysRevA.82.022310} {\bibfield  {journal}
  {\bibinfo  {journal} {Phys. Rev. A}\ }\textbf {\bibinfo {volume} {82}},\
  \bibinfo {pages} {022310} (\bibinfo {year} {2010})}\BibitemShut {NoStop}%
\bibitem [{\citenamefont {Dajczgewand}\ \emph {et~al.}(2014)\citenamefont
  {Dajczgewand}, \citenamefont {Gou\"{e}t}, \citenamefont {Louchet-Chauvet},\
  and\ \citenamefont {Chaneli\`{e}re}}]{Dajczgewand:14}%
  \BibitemOpen
  \bibfield  {author} {\bibinfo {author} {\bibfnamefont {J.}~\bibnamefont
  {Dajczgewand}}, \bibinfo {author} {\bibfnamefont {J.-L.~L.}\ \bibnamefont
  {Gou\"{e}t}}, \bibinfo {author} {\bibfnamefont {A.}~\bibnamefont
  {Louchet-Chauvet}}, \ and\ \bibinfo {author} {\bibfnamefont {T.}~\bibnamefont
  {Chaneli\`{e}re}},\ }\href {\doibase 10.1364/OL.39.002711} {\bibfield
  {journal} {\bibinfo  {journal} {Opt. Lett.}\ }\textbf {\bibinfo {volume}
  {39}},\ \bibinfo {pages} {2711} (\bibinfo {year} {2014})}\BibitemShut
  {NoStop}%
\bibitem [{\citenamefont {Moiseev}(2013)}]{SMoiseev2013}%
  \BibitemOpen
  \bibfield  {author} {\bibinfo {author} {\bibfnamefont {S.~A.}\ \bibnamefont
  {Moiseev}},\ }\href {\doibase 10.1103/PhysRevA.88.012304} {\bibfield
  {journal} {\bibinfo  {journal} {Phys. Rev. A}\ }\textbf {\bibinfo {volume}
  {88}},\ \bibinfo {pages} {012304} (\bibinfo {year} {2013})}\BibitemShut
  {NoStop}%
\bibitem [{\citenamefont {Hsiao}\ \emph {et~al.}(2018)\citenamefont {Hsiao},
  \citenamefont {Tsai}, \citenamefont {Chen}, \citenamefont {Lin},
  \citenamefont {Hung}, \citenamefont {Lee}, \citenamefont {Chen},
  \citenamefont {Chen}, \citenamefont {Yu},\ and\ \citenamefont
  {Chen}}]{Hsiao2018}%
  \BibitemOpen
  \bibfield  {author} {\bibinfo {author} {\bibfnamefont {Y.-F.}\ \bibnamefont
  {Hsiao}}, \bibinfo {author} {\bibfnamefont {P.-J.}\ \bibnamefont {Tsai}},
  \bibinfo {author} {\bibfnamefont {H.-S.}\ \bibnamefont {Chen}}, \bibinfo
  {author} {\bibfnamefont {S.-X.}\ \bibnamefont {Lin}}, \bibinfo {author}
  {\bibfnamefont {C.-C.}\ \bibnamefont {Hung}}, \bibinfo {author}
  {\bibfnamefont {C.-H.}\ \bibnamefont {Lee}}, \bibinfo {author} {\bibfnamefont
  {Y.-H.}\ \bibnamefont {Chen}}, \bibinfo {author} {\bibfnamefont {Y.-F.}\
  \bibnamefont {Chen}}, \bibinfo {author} {\bibfnamefont {I.~A.}\ \bibnamefont
  {Yu}}, \ and\ \bibinfo {author} {\bibfnamefont {Y.-C.}\ \bibnamefont
  {Chen}},\ }\href {\doibase 10.1103/PhysRevLett.120.183602} {\bibfield
  {journal} {\bibinfo  {journal} {Phys. Rev. Lett.}\ }\textbf {\bibinfo
  {volume} {120}},\ \bibinfo {pages} {183602} (\bibinfo {year}
  {2018})}\BibitemShut {NoStop}%
\bibitem [{\citenamefont {Yang}\ \emph {et~al.}(2016)\citenamefont {Yang},
  \citenamefont {Wang}, \citenamefont {Bao},\ and\ \citenamefont
  {Pan}}]{Yang2016}%
  \BibitemOpen
  \bibfield  {author} {\bibinfo {author} {\bibfnamefont {S.-J.}\ \bibnamefont
  {Yang}}, \bibinfo {author} {\bibfnamefont {X.-J.}\ \bibnamefont {Wang}},
  \bibinfo {author} {\bibfnamefont {X.-H.}\ \bibnamefont {Bao}}, \ and\
  \bibinfo {author} {\bibfnamefont {J.-W.}\ \bibnamefont {Pan}},\ }\href
  {https://doi.org/10.1038/nphoton.2016.51} {\bibfield  {journal} {\bibinfo
  {journal} {Nat. Photonics}\ }\textbf {\bibinfo {volume} {10}},\ \bibinfo
  {pages} {381} (\bibinfo {year} {2016})}\BibitemShut {NoStop}%
\bibitem [{\citenamefont {Radnaev}\ \emph {et~al.}(2010)\citenamefont
  {Radnaev}, \citenamefont {Dudin}, \citenamefont {Zhao}, \citenamefont {Jen},
  \citenamefont {Jenkins}, \citenamefont {Kuzmich},\ and\ \citenamefont
  {Kennedy}}]{Radnaev2010}%
  \BibitemOpen
  \bibfield  {author} {\bibinfo {author} {\bibfnamefont {A.~G.}\ \bibnamefont
  {Radnaev}}, \bibinfo {author} {\bibfnamefont {Y.~O.}\ \bibnamefont {Dudin}},
  \bibinfo {author} {\bibfnamefont {R.}~\bibnamefont {Zhao}}, \bibinfo {author}
  {\bibfnamefont {H.~H.}\ \bibnamefont {Jen}}, \bibinfo {author} {\bibfnamefont
  {S.~D.}\ \bibnamefont {Jenkins}}, \bibinfo {author} {\bibfnamefont
  {A.}~\bibnamefont {Kuzmich}}, \ and\ \bibinfo {author} {\bibfnamefont
  {T.~A.~B.}\ \bibnamefont {Kennedy}},\ }\href
  {https://doi.org/10.1038/nphys1773} {\bibfield  {journal} {\bibinfo
  {journal} {Nat. Phys.}\ }\textbf {\bibinfo {volume} {6}},\ \bibinfo {pages}
  {894} (\bibinfo {year} {2010})}\BibitemShut {NoStop}%
\bibitem [{\citenamefont {Craiciu}\ \emph {et~al.}(2019)\citenamefont
  {Craiciu}, \citenamefont {Lei}, \citenamefont {Rochman}, \citenamefont
  {Kindem}, \citenamefont {Bartholomew}, \citenamefont {Miyazono},
  \citenamefont {Zhong}, \citenamefont {Sinclair},\ and\ \citenamefont
  {Faraon}}]{Craiciu2019}%
  \BibitemOpen
  \bibfield  {author} {\bibinfo {author} {\bibfnamefont {I.}~\bibnamefont
  {Craiciu}}, \bibinfo {author} {\bibfnamefont {M.}~\bibnamefont {Lei}},
  \bibinfo {author} {\bibfnamefont {J.}~\bibnamefont {Rochman}}, \bibinfo
  {author} {\bibfnamefont {J.~M.}\ \bibnamefont {Kindem}}, \bibinfo {author}
  {\bibfnamefont {J.~G.}\ \bibnamefont {Bartholomew}}, \bibinfo {author}
  {\bibfnamefont {E.}~\bibnamefont {Miyazono}}, \bibinfo {author}
  {\bibfnamefont {T.}~\bibnamefont {Zhong}}, \bibinfo {author} {\bibfnamefont
  {N.}~\bibnamefont {Sinclair}}, \ and\ \bibinfo {author} {\bibfnamefont
  {A.}~\bibnamefont {Faraon}},\ }\href {\doibase
  10.1103/PhysRevApplied.12.024062} {\bibfield  {journal} {\bibinfo  {journal}
  {Phys. Rev. Appl.}\ }\textbf {\bibinfo {volume} {12}},\ \bibinfo {pages}
  {024062} (\bibinfo {year} {2019})}\BibitemShut {NoStop}%
\bibitem [{\citenamefont {Sparkes}\ \emph {et~al.}(2013)\citenamefont
  {Sparkes}, \citenamefont {Bernu}, \citenamefont {Hosseini}, \citenamefont
  {Geng}, \citenamefont {Glorieux}, \citenamefont {Altin}, \citenamefont {Lam},
  \citenamefont {Robins},\ and\ \citenamefont {Buchler}}]{Sparkes2013}%
  \BibitemOpen
  \bibfield  {author} {\bibinfo {author} {\bibfnamefont {B.~M.}\ \bibnamefont
  {Sparkes}}, \bibinfo {author} {\bibfnamefont {J.}~\bibnamefont {Bernu}},
  \bibinfo {author} {\bibfnamefont {M.}~\bibnamefont {Hosseini}}, \bibinfo
  {author} {\bibfnamefont {J.}~\bibnamefont {Geng}}, \bibinfo {author}
  {\bibfnamefont {Q.}~\bibnamefont {Glorieux}}, \bibinfo {author}
  {\bibfnamefont {P.~A.}\ \bibnamefont {Altin}}, \bibinfo {author}
  {\bibfnamefont {P.~K.}\ \bibnamefont {Lam}}, \bibinfo {author} {\bibfnamefont
  {N.~P.}\ \bibnamefont {Robins}}, \ and\ \bibinfo {author} {\bibfnamefont
  {B.~C.}\ \bibnamefont {Buchler}},\ }\href
  {http://stacks.iop.org/1367-2630/15/i=8/a=085027} {\bibfield  {journal}
  {\bibinfo  {journal} {New J. Phys.}\ }\textbf {\bibinfo {volume} {15}},\
  \bibinfo {pages} {085027} (\bibinfo {year} {2013})}\BibitemShut {NoStop}%
\bibitem [{\citenamefont {Manzoni}\ \emph {et~al.}(2018)\citenamefont
  {Manzoni}, \citenamefont {Moreno-Cardoner}, \citenamefont {Asenjo-Garcia},
  \citenamefont {Porto}, \citenamefont {Gorshkov},\ and\ \citenamefont
  {Chang}}]{ManzoniNJP2018}%
  \BibitemOpen
  \bibfield  {author} {\bibinfo {author} {\bibfnamefont {M.~T.}\ \bibnamefont
  {Manzoni}}, \bibinfo {author} {\bibfnamefont {M.}~\bibnamefont
  {Moreno-Cardoner}}, \bibinfo {author} {\bibfnamefont {A.}~\bibnamefont
  {Asenjo-Garcia}}, \bibinfo {author} {\bibfnamefont {J.~V.}\ \bibnamefont
  {Porto}}, \bibinfo {author} {\bibfnamefont {A.~V.}\ \bibnamefont {Gorshkov}},
  \ and\ \bibinfo {author} {\bibfnamefont {D.~E.}\ \bibnamefont {Chang}},\
  }\href {http://stacks.iop.org/1367-2630/20/i=8/a=083048} {\bibfield
  {journal} {\bibinfo  {journal} {New J. Phys.}\ }\textbf {\bibinfo {volume}
  {20}},\ \bibinfo {pages} {083048} (\bibinfo {year} {2018})}\BibitemShut
  {NoStop}%
\bibitem [{\citenamefont {Gorshkov}\ \emph
  {et~al.}(2007{\natexlab{a}})\citenamefont {Gorshkov}, \citenamefont
  {Andr\'e}, \citenamefont {Lukin},\ and\ \citenamefont
  {S\o{}rensen}}]{GorshkovPRA12007}%
  \BibitemOpen
  \bibfield  {author} {\bibinfo {author} {\bibfnamefont {A.~V.}\ \bibnamefont
  {Gorshkov}}, \bibinfo {author} {\bibfnamefont {A.}~\bibnamefont {Andr\'e}},
  \bibinfo {author} {\bibfnamefont {M.~D.}\ \bibnamefont {Lukin}}, \ and\
  \bibinfo {author} {\bibfnamefont {A.~S.}\ \bibnamefont {S\o{}rensen}},\
  }\href {\doibase 10.1103/PhysRevA.76.033804} {\bibfield  {journal} {\bibinfo
  {journal} {Phys. Rev. A}\ }\textbf {\bibinfo {volume} {76}},\ \bibinfo
  {pages} {033804} (\bibinfo {year} {2007}{\natexlab{a}})}\BibitemShut
  {NoStop}%
\bibitem [{\citenamefont {Moiseev}\ \emph {et~al.}(2010)\citenamefont
  {Moiseev}, \citenamefont {Andrianov},\ and\ \citenamefont
  {Gubaidullin}}]{Moiseev2010b}%
  \BibitemOpen
  \bibfield  {author} {\bibinfo {author} {\bibfnamefont {S.~A.}\ \bibnamefont
  {Moiseev}}, \bibinfo {author} {\bibfnamefont {S.~N.}\ \bibnamefont
  {Andrianov}}, \ and\ \bibinfo {author} {\bibfnamefont {F.~F.}\ \bibnamefont
  {Gubaidullin}},\ }\href {\doibase 10.1103/PhysRevA.82.022311} {\bibfield
  {journal} {\bibinfo  {journal} {Phys. Rev. A}\ }\textbf {\bibinfo {volume}
  {82}},\ \bibinfo {pages} {022311} (\bibinfo {year} {2010})}\BibitemShut
  {NoStop}%
\bibitem [{\citenamefont {Sabooni}\ \emph {et~al.}(2013)\citenamefont
  {Sabooni}, \citenamefont {Li}, \citenamefont {Rippe}, \citenamefont {Mohan},\
  and\ \citenamefont {Kr\"oll}}]{SabooniPRL2013}%
  \BibitemOpen
  \bibfield  {author} {\bibinfo {author} {\bibfnamefont {M.}~\bibnamefont
  {Sabooni}}, \bibinfo {author} {\bibfnamefont {Q.}~\bibnamefont {Li}},
  \bibinfo {author} {\bibfnamefont {L.}~\bibnamefont {Rippe}}, \bibinfo
  {author} {\bibfnamefont {R.~K.}\ \bibnamefont {Mohan}}, \ and\ \bibinfo
  {author} {\bibfnamefont {S.}~\bibnamefont {Kr\"oll}},\ }\href {\doibase
  10.1103/PhysRevLett.111.183602} {\bibfield  {journal} {\bibinfo  {journal}
  {Phys. Rev. Lett.}\ }\textbf {\bibinfo {volume} {111}},\ \bibinfo {pages}
  {183602} (\bibinfo {year} {2013})}\BibitemShut {NoStop}%
\bibitem [{\citenamefont {Wicht}\ \emph {et~al.}(1997)\citenamefont {Wicht},
  \citenamefont {Danzmann}, \citenamefont {Fleischhauer}, \citenamefont
  {Scully}, \citenamefont {Müller},\ and\ \citenamefont
  {Rinkleff}}]{WICHT1997431}%
  \BibitemOpen
  \bibfield  {author} {\bibinfo {author} {\bibfnamefont {A.}~\bibnamefont
  {Wicht}}, \bibinfo {author} {\bibfnamefont {K.}~\bibnamefont {Danzmann}},
  \bibinfo {author} {\bibfnamefont {M.}~\bibnamefont {Fleischhauer}}, \bibinfo
  {author} {\bibfnamefont {M.}~\bibnamefont {Scully}}, \bibinfo {author}
  {\bibfnamefont {G.}~\bibnamefont {Müller}}, \ and\ \bibinfo {author}
  {\bibfnamefont {R.-H.}\ \bibnamefont {Rinkleff}},\ }\href {\doibase
  https://doi.org/10.1016/S0030-4018(96)00579-2} {\bibfield  {journal}
  {\bibinfo  {journal} {Opt. Commun.}\ }\textbf {\bibinfo {volume} {134}},\
  \bibinfo {pages} {431} (\bibinfo {year} {1997})}\BibitemShut {NoStop}%
\bibitem [{\citenamefont {Xu}\ \emph {et~al.}(2012)\citenamefont {Xu},
  \citenamefont {Al-Amri}, \citenamefont {Yang}, \citenamefont {Zhu},\ and\
  \citenamefont {Zubairy}}]{Xu2012}%
  \BibitemOpen
  \bibfield  {author} {\bibinfo {author} {\bibfnamefont {J.}~\bibnamefont
  {Xu}}, \bibinfo {author} {\bibfnamefont {M.}~\bibnamefont {Al-Amri}},
  \bibinfo {author} {\bibfnamefont {Y.}~\bibnamefont {Yang}}, \bibinfo {author}
  {\bibfnamefont {S.-Y.}\ \bibnamefont {Zhu}}, \ and\ \bibinfo {author}
  {\bibfnamefont {M.~S.}\ \bibnamefont {Zubairy}},\ }\href {\doibase
  10.1103/PhysRevA.86.033828} {\bibfield  {journal} {\bibinfo  {journal} {Phys.
  Rev. A}\ }\textbf {\bibinfo {volume} {86}},\ \bibinfo {pages} {033828}
  (\bibinfo {year} {2012})}\BibitemShut {NoStop}%
\bibitem [{\citenamefont {Li}\ \emph {et~al.}(2016)\citenamefont {Li},
  \citenamefont {Xu}, \citenamefont {Song}, \citenamefont {Zhu}, \citenamefont
  {Xie}, \citenamefont {Yang}, \citenamefont {Zubairy},\ and\ \citenamefont
  {Zhu}}]{Li2016}%
  \BibitemOpen
  \bibfield  {author} {\bibinfo {author} {\bibfnamefont {N.}~\bibnamefont
  {Li}}, \bibinfo {author} {\bibfnamefont {J.}~\bibnamefont {Xu}}, \bibinfo
  {author} {\bibfnamefont {G.}~\bibnamefont {Song}}, \bibinfo {author}
  {\bibfnamefont {C.}~\bibnamefont {Zhu}}, \bibinfo {author} {\bibfnamefont
  {S.}~\bibnamefont {Xie}}, \bibinfo {author} {\bibfnamefont {Y.}~\bibnamefont
  {Yang}}, \bibinfo {author} {\bibfnamefont {M.~S.}\ \bibnamefont {Zubairy}}, \
  and\ \bibinfo {author} {\bibfnamefont {S.-Y.}\ \bibnamefont {Zhu}},\ }\href
  {\doibase 10.1103/PhysRevA.93.043819} {\bibfield  {journal} {\bibinfo
  {journal} {Phys. Rev. A}\ }\textbf {\bibinfo {volume} {93}},\ \bibinfo
  {pages} {043819} (\bibinfo {year} {2016})}\BibitemShut {NoStop}%
\bibitem [{\citenamefont {Pati}\ \emph {et~al.}(2007)\citenamefont {Pati},
  \citenamefont {Salit}, \citenamefont {Salit},\ and\ \citenamefont
  {Shahriar}}]{Pati2007}%
  \BibitemOpen
  \bibfield  {author} {\bibinfo {author} {\bibfnamefont {G.~S.}\ \bibnamefont
  {Pati}}, \bibinfo {author} {\bibfnamefont {M.}~\bibnamefont {Salit}},
  \bibinfo {author} {\bibfnamefont {K.}~\bibnamefont {Salit}}, \ and\ \bibinfo
  {author} {\bibfnamefont {M.~S.}\ \bibnamefont {Shahriar}},\ }\href {\doibase
  10.1103/PhysRevLett.99.133601} {\bibfield  {journal} {\bibinfo  {journal}
  {Phys. Rev. Lett.}\ }\textbf {\bibinfo {volume} {99}},\ \bibinfo {pages}
  {133601} (\bibinfo {year} {2007})}\BibitemShut {NoStop}%
\bibitem [{\citenamefont {Wu}\ and\ \citenamefont {Xiao}(2008)}]{Wu2008}%
  \BibitemOpen
  \bibfield  {author} {\bibinfo {author} {\bibfnamefont {H.}~\bibnamefont
  {Wu}}\ and\ \bibinfo {author} {\bibfnamefont {M.}~\bibnamefont {Xiao}},\
  }\href {\doibase 10.1103/PhysRevA.77.031801} {\bibfield  {journal} {\bibinfo
  {journal} {Phys. Rev. A}\ }\textbf {\bibinfo {volume} {77}},\ \bibinfo
  {pages} {031801(R)} (\bibinfo {year} {2008})}\BibitemShut {NoStop}%
\bibitem [{\citenamefont {Ma}\ \emph {et~al.}(2015)\citenamefont {Ma},
  \citenamefont {Miao}, \citenamefont {Zhao},\ and\ \citenamefont
  {Chen}}]{Ma2015}%
  \BibitemOpen
  \bibfield  {author} {\bibinfo {author} {\bibfnamefont {Y.}~\bibnamefont
  {Ma}}, \bibinfo {author} {\bibfnamefont {H.}~\bibnamefont {Miao}}, \bibinfo
  {author} {\bibfnamefont {C.}~\bibnamefont {Zhao}}, \ and\ \bibinfo {author}
  {\bibfnamefont {Y.}~\bibnamefont {Chen}},\ }\href {\doibase
  10.1103/PhysRevA.92.023807} {\bibfield  {journal} {\bibinfo  {journal} {Phys.
  Rev. A}\ }\textbf {\bibinfo {volume} {92}},\ \bibinfo {pages} {023807}
  (\bibinfo {year} {2015})}\BibitemShut {NoStop}%
\bibitem [{\citenamefont {Nilsson}\ \emph {et~al.}(2004)\citenamefont
  {Nilsson}, \citenamefont {Rippe}, \citenamefont {Kr\"oll}, \citenamefont
  {Klieber},\ and\ \citenamefont {Suter}}]{Nilsson2004}%
  \BibitemOpen
  \bibfield  {author} {\bibinfo {author} {\bibfnamefont {M.}~\bibnamefont
  {Nilsson}}, \bibinfo {author} {\bibfnamefont {L.}~\bibnamefont {Rippe}},
  \bibinfo {author} {\bibfnamefont {S.}~\bibnamefont {Kr\"oll}}, \bibinfo
  {author} {\bibfnamefont {R.}~\bibnamefont {Klieber}}, \ and\ \bibinfo
  {author} {\bibfnamefont {D.}~\bibnamefont {Suter}},\ }\href {\doibase
  10.1103/PhysRevB.70.214116} {\bibfield  {journal} {\bibinfo  {journal} {Phys.
  Rev. B}\ }\textbf {\bibinfo {volume} {70}},\ \bibinfo {pages} {214116}
  (\bibinfo {year} {2004})}\BibitemShut {NoStop}%
\bibitem [{\citenamefont {Amari}\ \emph {et~al.}(2010)\citenamefont {Amari},
  \citenamefont {Walther}, \citenamefont {Sabooni}, \citenamefont {Huang},
  \citenamefont {Kr{\"o}ll}, \citenamefont {Afzelius}, \citenamefont {Usmani},
  \citenamefont {Lauritzen}, \citenamefont {Sangouard}, \citenamefont
  {De~Riedmatten},\ and\ \citenamefont {Gisin}}]{amari2010towards}%
  \BibitemOpen
  \bibfield  {author} {\bibinfo {author} {\bibfnamefont {A.}~\bibnamefont
  {Amari}}, \bibinfo {author} {\bibfnamefont {A.}~\bibnamefont {Walther}},
  \bibinfo {author} {\bibfnamefont {M.}~\bibnamefont {Sabooni}}, \bibinfo
  {author} {\bibfnamefont {M.}~\bibnamefont {Huang}}, \bibinfo {author}
  {\bibfnamefont {S.}~\bibnamefont {Kr{\"o}ll}}, \bibinfo {author}
  {\bibfnamefont {M.}~\bibnamefont {Afzelius}}, \bibinfo {author}
  {\bibfnamefont {I.}~\bibnamefont {Usmani}}, \bibinfo {author} {\bibfnamefont
  {B.}~\bibnamefont {Lauritzen}}, \bibinfo {author} {\bibfnamefont
  {N.}~\bibnamefont {Sangouard}}, \bibinfo {author} {\bibfnamefont
  {H.}~\bibnamefont {De~Riedmatten}}, \ and\ \bibinfo {author} {\bibfnamefont
  {N.}~\bibnamefont {Gisin}},\ }\href@noop {} {\bibfield  {journal} {\bibinfo
  {journal} {J. Lumin.}\ }\textbf {\bibinfo {volume} {130}},\ \bibinfo {pages}
  {1579} (\bibinfo {year} {2010})}\BibitemShut {NoStop}%
\bibitem [{\citenamefont {Gardiner}\ and\ \citenamefont
  {Collett}(1985)}]{Gardiner1985}%
  \BibitemOpen
  \bibfield  {author} {\bibinfo {author} {\bibfnamefont {C.~W.}\ \bibnamefont
  {Gardiner}}\ and\ \bibinfo {author} {\bibfnamefont {M.~J.}\ \bibnamefont
  {Collett}},\ }\href {\doibase 10.1103/PhysRevA.31.3761} {\bibfield  {journal}
  {\bibinfo  {journal} {Phys. Rev. A}\ }\textbf {\bibinfo {volume} {31}},\
  \bibinfo {pages} {3761} (\bibinfo {year} {1985})}\BibitemShut {NoStop}%
\bibitem [{\citenamefont {Scully}\ and\ \citenamefont
  {Zubairy}(1997)}]{Scully1997}%
  \BibitemOpen
  \bibfield  {author} {\bibinfo {author} {\bibfnamefont {M.~O.}\ \bibnamefont
  {Scully}}\ and\ \bibinfo {author} {\bibfnamefont {S.~M.}\ \bibnamefont
  {Zubairy}},\ }\href@noop {} {\emph {\bibinfo {title} {Quantum Optics}}}\
  (\bibinfo  {publisher} {Cambridge Univ. Press},\ \bibinfo {address}
  {Cambridge},\ \bibinfo {year} {1997})\BibitemShut {NoStop}%
\bibitem [{\citenamefont {Lauk}\ \emph {et~al.}(2013)\citenamefont {Lauk},
  \citenamefont {O'Brien},\ and\ \citenamefont
  {Fleischhauer}}]{FleischhauerPRA2013}%
  \BibitemOpen
  \bibfield  {author} {\bibinfo {author} {\bibfnamefont {N.}~\bibnamefont
  {Lauk}}, \bibinfo {author} {\bibfnamefont {C.}~\bibnamefont {O'Brien}}, \
  and\ \bibinfo {author} {\bibfnamefont {M.}~\bibnamefont {Fleischhauer}},\
  }\href {\doibase 10.1103/PhysRevA.88.013823} {\bibfield  {journal} {\bibinfo
  {journal} {Phys. Rev. A}\ }\textbf {\bibinfo {volume} {88}},\ \bibinfo
  {pages} {013823} (\bibinfo {year} {2013})}\BibitemShut {NoStop}%
\bibitem [{\citenamefont {Lauk}(2016)}]{lauk2016quantum}%
  \BibitemOpen
  \bibfield  {author} {\bibinfo {author} {\bibfnamefont {N.}~\bibnamefont
  {Lauk}},\ }\emph {\bibinfo {title} {Quantum Networks for Photons: Nonlinear
  Effects in Quantum Memories, Quantum Interfaces and Single-photon Filter}},\
  \href@noop {} {Ph.D. thesis},\ \bibinfo  {school} {Technische Universit{\"a}t
  Kaiserslautern} (\bibinfo {year} {2016})\BibitemShut {NoStop}%
\bibitem [{\citenamefont {Gorshkov}\ \emph
  {et~al.}(2007{\natexlab{b}})\citenamefont {Gorshkov}, \citenamefont
  {Andr\'e}, \citenamefont {Lukin},\ and\ \citenamefont
  {S\o{}rensen}}]{GorshkovPRA2007}%
  \BibitemOpen
  \bibfield  {author} {\bibinfo {author} {\bibfnamefont {A.~V.}\ \bibnamefont
  {Gorshkov}}, \bibinfo {author} {\bibfnamefont {A.}~\bibnamefont {Andr\'e}},
  \bibinfo {author} {\bibfnamefont {M.~D.}\ \bibnamefont {Lukin}}, \ and\
  \bibinfo {author} {\bibfnamefont {A.~S.}\ \bibnamefont {S\o{}rensen}},\
  }\href {\doibase 10.1103/PhysRevA.76.033805} {\bibfield  {journal} {\bibinfo
  {journal} {Phys. Rev. A}\ }\textbf {\bibinfo {volume} {76}},\ \bibinfo
  {pages} {033805} (\bibinfo {year} {2007}{\natexlab{b}})}\BibitemShut
  {NoStop}%
\bibitem [{\citenamefont {Moiseev}\ and\ \citenamefont
  {Moiseev}(2013)}]{EMoiseev2013}%
  \BibitemOpen
  \bibfield  {author} {\bibinfo {author} {\bibfnamefont {E.~S.}\ \bibnamefont
  {Moiseev}}\ and\ \bibinfo {author} {\bibfnamefont {S.~A.}\ \bibnamefont
  {Moiseev}},\ }\href {http://stacks.iop.org/1367-2630/15/i=10/a=105005}
  {\bibfield  {journal} {\bibinfo  {journal} {New J. Phys.}\ }\textbf {\bibinfo
  {volume} {15}},\ \bibinfo {pages} {105005} (\bibinfo {year}
  {2013})}\BibitemShut {NoStop}%
\bibitem [{\citenamefont {Bonarota}\ \emph {et~al.}(2014)\citenamefont
  {Bonarota}, \citenamefont {Dajczgewand}, \citenamefont {Louchet-Chauvet},
  \citenamefont {Gouët},\ and\ \citenamefont
  {Chaneli{\`{e}}re}}]{Bonarota2014}%
  \BibitemOpen
  \bibfield  {author} {\bibinfo {author} {\bibfnamefont {M.}~\bibnamefont
  {Bonarota}}, \bibinfo {author} {\bibfnamefont {J.}~\bibnamefont
  {Dajczgewand}}, \bibinfo {author} {\bibfnamefont {A.}~\bibnamefont
  {Louchet-Chauvet}}, \bibinfo {author} {\bibfnamefont {J.-L.~L.}\ \bibnamefont
  {Gouët}}, \ and\ \bibinfo {author} {\bibfnamefont {T.}~\bibnamefont
  {Chaneli{\`{e}}re}},\ }\href {\doibase 10.1088/1054-660x/24/9/094003}
  {\bibfield  {journal} {\bibinfo  {journal} {Laser Physics}\ }\textbf
  {\bibinfo {volume} {24}},\ \bibinfo {pages} {094003} (\bibinfo {year}
  {2014})}\BibitemShut {NoStop}%
\bibitem [{\citenamefont {de~Riedmatten}\ \emph {et~al.}(2008)\citenamefont
  {de~Riedmatten}, \citenamefont {Afzelius}, \citenamefont {Staudt},
  \citenamefont {Simon},\ and\ \citenamefont {Gisin}}]{Riedmatten2008}%
  \BibitemOpen
  \bibfield  {author} {\bibinfo {author} {\bibfnamefont {H.}~\bibnamefont
  {de~Riedmatten}}, \bibinfo {author} {\bibfnamefont {M.}~\bibnamefont
  {Afzelius}}, \bibinfo {author} {\bibfnamefont {M.~U.}\ \bibnamefont
  {Staudt}}, \bibinfo {author} {\bibfnamefont {C.}~\bibnamefont {Simon}}, \
  and\ \bibinfo {author} {\bibfnamefont {N.}~\bibnamefont {Gisin}},\ }\href
  {\doibase 10.1038/nature07607} {\bibfield  {journal} {\bibinfo  {journal}
  {Nature}\ }\textbf {\bibinfo {volume} {456}},\ \bibinfo {pages} {773}
  (\bibinfo {year} {2008})}\BibitemShut {NoStop}%
\bibitem [{\citenamefont {Simon}\ \emph {et~al.}(2010)\citenamefont {Simon},
  \citenamefont {Afzelius}, \citenamefont {Appel}, \citenamefont {{Boyer De La
  Giroday}}, \citenamefont {Dewhurst}, \citenamefont {Gisin}, \citenamefont
  {Hu}, \citenamefont {Jelezko}, \citenamefont {Kr{\"{o}}ll}, \citenamefont
  {M{\"{u}}ller}, \citenamefont {Nunn}, \citenamefont {Polzik}, \citenamefont
  {Rarity}, \citenamefont {{De Riedmatten}}, \citenamefont {Rosenfeld},
  \citenamefont {Shields}, \citenamefont {Sk{\"{o}}ld}, \citenamefont
  {Stevenson}, \citenamefont {Thew}, \citenamefont {Walmsley}, \citenamefont
  {Weber}, \citenamefont {Weinfurter}, \citenamefont {Wrachtrup},\ and\
  \citenamefont {Young}}]{Simon2010a}%
  \BibitemOpen
  \bibfield  {author} {\bibinfo {author} {\bibfnamefont {C.}~\bibnamefont
  {Simon}}, \bibinfo {author} {\bibfnamefont {M.}~\bibnamefont {Afzelius}},
  \bibinfo {author} {\bibfnamefont {J.}~\bibnamefont {Appel}}, \bibinfo
  {author} {\bibfnamefont {A.}~\bibnamefont {{Boyer De La Giroday}}}, \bibinfo
  {author} {\bibfnamefont {S.~J.}\ \bibnamefont {Dewhurst}}, \bibinfo {author}
  {\bibfnamefont {N.}~\bibnamefont {Gisin}}, \bibinfo {author} {\bibfnamefont
  {C.~Y.}\ \bibnamefont {Hu}}, \bibinfo {author} {\bibfnamefont
  {F.}~\bibnamefont {Jelezko}}, \bibinfo {author} {\bibfnamefont
  {S.}~\bibnamefont {Kr{\"{o}}ll}}, \bibinfo {author} {\bibfnamefont {J.~H.}\
  \bibnamefont {M{\"{u}}ller}}, \bibinfo {author} {\bibfnamefont
  {J.}~\bibnamefont {Nunn}}, \bibinfo {author} {\bibfnamefont {E.~S.}\
  \bibnamefont {Polzik}}, \bibinfo {author} {\bibfnamefont {J.~G.}\
  \bibnamefont {Rarity}}, \bibinfo {author} {\bibfnamefont {H.}~\bibnamefont
  {{De Riedmatten}}}, \bibinfo {author} {\bibfnamefont {W.}~\bibnamefont
  {Rosenfeld}}, \bibinfo {author} {\bibfnamefont {A.~J.}\ \bibnamefont
  {Shields}}, \bibinfo {author} {\bibfnamefont {N.}~\bibnamefont
  {Sk{\"{o}}ld}}, \bibinfo {author} {\bibfnamefont {R.~M.}\ \bibnamefont
  {Stevenson}}, \bibinfo {author} {\bibfnamefont {R.}~\bibnamefont {Thew}},
  \bibinfo {author} {\bibfnamefont {I.~a.}\ \bibnamefont {Walmsley}}, \bibinfo
  {author} {\bibfnamefont {M.~C.}\ \bibnamefont {Weber}}, \bibinfo {author}
  {\bibfnamefont {H.}~\bibnamefont {Weinfurter}}, \bibinfo {author}
  {\bibfnamefont {J.}~\bibnamefont {Wrachtrup}}, \ and\ \bibinfo {author}
  {\bibfnamefont {R.~J.}\ \bibnamefont {Young}},\ }\href {\doibase
  10.1140/epjd/e2010-00103-y} {\bibfield  {journal} {\bibinfo  {journal} {Eur.
  Phys. J. D}\ }\textbf {\bibinfo {volume} {58}},\ \bibinfo {pages} {1}
  (\bibinfo {year} {2010})}\BibitemShut {NoStop}%
\bibitem [{\citenamefont {Moiseev}\ and\ \citenamefont
  {Gou\"{e}t}(2012)}]{Moiseev2012}%
  \BibitemOpen
  \bibfield  {author} {\bibinfo {author} {\bibfnamefont {S.~A.}\ \bibnamefont
  {Moiseev}}\ and\ \bibinfo {author} {\bibfnamefont {J.-L.~L.}\ \bibnamefont
  {Gou\"{e}t}},\ }\href {http://stacks.iop.org/0953-4075/45/i=12/a=124003}
  {\bibfield  {journal} {\bibinfo  {journal} {Journal of Physics B: Atomic,
  Molecular and Optical Physics}\ }\textbf {\bibinfo {volume} {45}},\ \bibinfo
  {pages} {124003} (\bibinfo {year} {2012})}\BibitemShut {NoStop}%
\bibitem [{\citenamefont {Nunn}\ \emph {et~al.}(2017)\citenamefont {Nunn},
  \citenamefont {Munns}, \citenamefont {Thomas}, \citenamefont {Kaczmarek},
  \citenamefont {Qiu}, \citenamefont {Feizpour}, \citenamefont {Poem},
  \citenamefont {Brecht}, \citenamefont {Saunders}, \citenamefont {Ledingham},
  \citenamefont {Reddy}, \citenamefont {Raymer},\ and\ \citenamefont
  {Walmsley}}]{Nunn2017}%
  \BibitemOpen
  \bibfield  {author} {\bibinfo {author} {\bibfnamefont {J.}~\bibnamefont
  {Nunn}}, \bibinfo {author} {\bibfnamefont {J.~H.~D.}\ \bibnamefont {Munns}},
  \bibinfo {author} {\bibfnamefont {S.}~\bibnamefont {Thomas}}, \bibinfo
  {author} {\bibfnamefont {K.~T.}\ \bibnamefont {Kaczmarek}}, \bibinfo {author}
  {\bibfnamefont {C.}~\bibnamefont {Qiu}}, \bibinfo {author} {\bibfnamefont
  {A.}~\bibnamefont {Feizpour}}, \bibinfo {author} {\bibfnamefont
  {E.}~\bibnamefont {Poem}}, \bibinfo {author} {\bibfnamefont {B.}~\bibnamefont
  {Brecht}}, \bibinfo {author} {\bibfnamefont {D.~J.}\ \bibnamefont
  {Saunders}}, \bibinfo {author} {\bibfnamefont {P.~M.}\ \bibnamefont
  {Ledingham}}, \bibinfo {author} {\bibfnamefont {D.~V.}\ \bibnamefont
  {Reddy}}, \bibinfo {author} {\bibfnamefont {M.~G.}\ \bibnamefont {Raymer}}, \
  and\ \bibinfo {author} {\bibfnamefont {I.~A.}\ \bibnamefont {Walmsley}},\
  }\href {\doibase 10.1103/PhysRevA.96.012338} {\bibfield  {journal} {\bibinfo
  {journal} {Phys. Rev. A}\ }\textbf {\bibinfo {volume} {96}},\ \bibinfo
  {pages} {012338} (\bibinfo {year} {2017})}\BibitemShut {NoStop}%
\bibitem [{\citenamefont {Muschik}\ \emph {et~al.}(2011)\citenamefont
  {Muschik}, \citenamefont {Polzik},\ and\ \citenamefont
  {Cirac}}]{PolzikCirac2011}%
  \BibitemOpen
  \bibfield  {author} {\bibinfo {author} {\bibfnamefont {C.~A.}\ \bibnamefont
  {Muschik}}, \bibinfo {author} {\bibfnamefont {E.~S.}\ \bibnamefont {Polzik}},
  \ and\ \bibinfo {author} {\bibfnamefont {J.~I.}\ \bibnamefont {Cirac}},\
  }\href {\doibase 10.1103/PhysRevA.83.052312} {\bibfield  {journal} {\bibinfo
  {journal} {Phys. Rev. A}\ }\textbf {\bibinfo {volume} {83}},\ \bibinfo
  {pages} {052312} (\bibinfo {year} {2011})}\BibitemShut {NoStop}%
\bibitem [{\citenamefont {Moiseev}\ \emph {et~al.}(2020)\citenamefont
  {Moiseev}, \citenamefont {Tashchilina}, \citenamefont {Moiseev},\ and\
  \citenamefont {Lvovsky}}]{Moiseev2020}%
  \BibitemOpen
  \bibfield  {author} {\bibinfo {author} {\bibfnamefont {E.~S.}\ \bibnamefont
  {Moiseev}}, \bibinfo {author} {\bibfnamefont {A.}~\bibnamefont
  {Tashchilina}}, \bibinfo {author} {\bibfnamefont {S.~A.}\ \bibnamefont
  {Moiseev}}, \ and\ \bibinfo {author} {\bibfnamefont {A.~I.}\ \bibnamefont
  {Lvovsky}},\ }\href {\doibase 10.1088/1367-2630/ab5fac} {\bibfield  {journal}
  {\bibinfo  {journal} {New J. Phys.}\ }\textbf {\bibinfo {volume} {22}},\
  \bibinfo {pages} {013014} (\bibinfo {year} {2020})}\BibitemShut {NoStop}%
\bibitem [{\citenamefont {Lobino}\ \emph {et~al.}(2009)\citenamefont {Lobino},
  \citenamefont {Kupchak}, \citenamefont {Figueroa},\ and\ \citenamefont
  {Lvovsky}}]{Lobino2009}%
  \BibitemOpen
  \bibfield  {author} {\bibinfo {author} {\bibfnamefont {M.}~\bibnamefont
  {Lobino}}, \bibinfo {author} {\bibfnamefont {C.}~\bibnamefont {Kupchak}},
  \bibinfo {author} {\bibfnamefont {E.}~\bibnamefont {Figueroa}}, \ and\
  \bibinfo {author} {\bibfnamefont {A.~I.}\ \bibnamefont {Lvovsky}},\ }\href
  {\doibase 10.1103/PhysRevLett.102.203601} {\bibfield  {journal} {\bibinfo
  {journal} {Phys. Rev. Lett.}\ }\textbf {\bibinfo {volume} {102}},\ \bibinfo
  {pages} {203601} (\bibinfo {year} {2009})}\BibitemShut {NoStop}%
\bibitem [{\citenamefont {Sangouard}\ \emph {et~al.}(2011)\citenamefont
  {Sangouard}, \citenamefont {Simon}, \citenamefont {de~Riedmatten},\ and\
  \citenamefont {Gisin}}]{Gisin2011}%
  \BibitemOpen
  \bibfield  {author} {\bibinfo {author} {\bibfnamefont {N.}~\bibnamefont
  {Sangouard}}, \bibinfo {author} {\bibfnamefont {C.}~\bibnamefont {Simon}},
  \bibinfo {author} {\bibfnamefont {H.}~\bibnamefont {de~Riedmatten}}, \ and\
  \bibinfo {author} {\bibfnamefont {N.}~\bibnamefont {Gisin}},\ }\href
  {\doibase 10.1103/RevModPhys.83.33} {\bibfield  {journal} {\bibinfo
  {journal} {Rev. Mod. Phys.}\ }\textbf {\bibinfo {volume} {83}},\ \bibinfo
  {pages} {33} (\bibinfo {year} {2011})}\BibitemShut {NoStop}%
\bibitem [{\citenamefont {Askarani}\ \emph {et~al.}(2019)\citenamefont
  {Askarani}, \citenamefont {Puigibert}, \citenamefont {Lutz}, \citenamefont
  {Verma}, \citenamefont {Shaw}, \citenamefont {Nam}, \citenamefont {Sinclair},
  \citenamefont {Oblak},\ and\ \citenamefont {Tittel}}]{Tittel2019PRAppl}%
  \BibitemOpen
  \bibfield  {author} {\bibinfo {author} {\bibfnamefont {M.~F.}\ \bibnamefont
  {Askarani}}, \bibinfo {author} {\bibfnamefont {M.}~\bibnamefont {Puigibert}},
  \bibinfo {author} {\bibfnamefont {T.}~\bibnamefont {Lutz}}, \bibinfo {author}
  {\bibfnamefont {V.~B.}\ \bibnamefont {Verma}}, \bibinfo {author}
  {\bibfnamefont {M.~D.}\ \bibnamefont {Shaw}}, \bibinfo {author}
  {\bibfnamefont {S.~W.}\ \bibnamefont {Nam}}, \bibinfo {author} {\bibfnamefont
  {N.}~\bibnamefont {Sinclair}}, \bibinfo {author} {\bibfnamefont
  {D.}~\bibnamefont {Oblak}}, \ and\ \bibinfo {author} {\bibfnamefont
  {W.}~\bibnamefont {Tittel}},\ }\href {\doibase
  10.1103/PhysRevApplied.11.054056} {\bibfield  {journal} {\bibinfo  {journal}
  {Phys. Rev. Appl.}\ }\textbf {\bibinfo {volume} {11}},\ \bibinfo {pages}
  {054056} (\bibinfo {year} {2019})}\BibitemShut {NoStop}%
\bibitem [{\citenamefont {Ran{\v{c}}i{\'{c}}}\ \emph
  {et~al.}(2018)\citenamefont {Ran{\v{c}}i{\'{c}}}, \citenamefont {Hedges},
  \citenamefont {Ahlefeldt},\ and\ \citenamefont {Sellars}}]{2017-Rancic-NP}%
  \BibitemOpen
  \bibfield  {author} {\bibinfo {author} {\bibfnamefont {M.}~\bibnamefont
  {Ran{\v{c}}i{\'{c}}}}, \bibinfo {author} {\bibfnamefont {M.~P.}\ \bibnamefont
  {Hedges}}, \bibinfo {author} {\bibfnamefont {R.~L.}\ \bibnamefont
  {Ahlefeldt}}, \ and\ \bibinfo {author} {\bibfnamefont {M.~J.}\ \bibnamefont
  {Sellars}},\ }\href {\doibase 10.1038/nphys4254} {\bibfield  {journal}
  {\bibinfo  {journal} {Nat. Phys.}\ }\textbf {\bibinfo {volume} {14}},\
  \bibinfo {pages} {50} (\bibinfo {year} {2018})}\BibitemShut {NoStop}%
\bibitem [{\citenamefont {Jobez}\ \emph {et~al.}(2014)\citenamefont {Jobez},
  \citenamefont {Usmani}, \citenamefont {Timoney}, \citenamefont {Laplane},
  \citenamefont {Gisin},\ and\ \citenamefont {Afzelius}}]{Jobez2014}%
  \BibitemOpen
  \bibfield  {author} {\bibinfo {author} {\bibfnamefont {P.}~\bibnamefont
  {Jobez}}, \bibinfo {author} {\bibfnamefont {I.}~\bibnamefont {Usmani}},
  \bibinfo {author} {\bibfnamefont {N.}~\bibnamefont {Timoney}}, \bibinfo
  {author} {\bibfnamefont {C.}~\bibnamefont {Laplane}}, \bibinfo {author}
  {\bibfnamefont {N.}~\bibnamefont {Gisin}}, \ and\ \bibinfo {author}
  {\bibfnamefont {M.}~\bibnamefont {Afzelius}},\ }\href {\doibase
  10.1088/1367-2630/16/8/083005} {\bibfield  {journal} {\bibinfo  {journal}
  {New J. Phys.}\ }\textbf {\bibinfo {volume} {16}},\ \bibinfo {pages} {83005}
  (\bibinfo {year} {2014})}\BibitemShut {NoStop}%
\bibitem [{\citenamefont {Minnegaliev}\ \emph {et~al.}(2017)\citenamefont
  {Minnegaliev}, \citenamefont {Baibekov}, \citenamefont {Gerasimov},
  \citenamefont {Moiseev}, \citenamefont {Smirnov},\ and\ \citenamefont
  {Urmancheev}}]{2017-Minnegaliev-QE}%
  \BibitemOpen
  \bibfield  {author} {\bibinfo {author} {\bibfnamefont {M.~M.}\ \bibnamefont
  {Minnegaliev}}, \bibinfo {author} {\bibfnamefont {E.~I.}\ \bibnamefont
  {Baibekov}}, \bibinfo {author} {\bibfnamefont {K.~I.}\ \bibnamefont
  {Gerasimov}}, \bibinfo {author} {\bibfnamefont {S.~A.}\ \bibnamefont
  {Moiseev}}, \bibinfo {author} {\bibfnamefont {M.~A.}\ \bibnamefont
  {Smirnov}}, \ and\ \bibinfo {author} {\bibfnamefont {R.~V.}\ \bibnamefont
  {Urmancheev}},\ }\href {\doibase 10.1070/QEL16470} {\bibfield  {journal}
  {\bibinfo  {journal} {Quantum Electron.}\ }\textbf {\bibinfo {volume} {47}},\
  \bibinfo {pages} {778} (\bibinfo {year} {2017})}\BibitemShut {NoStop}%
\bibitem [{\citenamefont {Goto}\ \emph {et~al.}(2010)\citenamefont {Goto},
  \citenamefont {Nakamura},\ and\ \citenamefont {Ichimura}}]{Goto2010}%
  \BibitemOpen
  \bibfield  {author} {\bibinfo {author} {\bibfnamefont {H.}~\bibnamefont
  {Goto}}, \bibinfo {author} {\bibfnamefont {S.}~\bibnamefont {Nakamura}}, \
  and\ \bibinfo {author} {\bibfnamefont {K.}~\bibnamefont {Ichimura}},\ }\href
  {\doibase 10.1364/oe.18.023763} {\bibfield  {journal} {\bibinfo  {journal}
  {Opt. Express}\ }\textbf {\bibinfo {volume} {18}},\ \bibinfo {pages} {23763}
  (\bibinfo {year} {2010})}\BibitemShut {NoStop}%
\bibitem [{\citenamefont {McAuslan}\ \emph {et~al.}(2011)\citenamefont
  {McAuslan}, \citenamefont {Korystov},\ and\ \citenamefont
  {Longdell}}]{McAuslan2011PRA}%
  \BibitemOpen
  \bibfield  {author} {\bibinfo {author} {\bibfnamefont {D.~L.}\ \bibnamefont
  {McAuslan}}, \bibinfo {author} {\bibfnamefont {D.}~\bibnamefont {Korystov}},
  \ and\ \bibinfo {author} {\bibfnamefont {J.~J.}\ \bibnamefont {Longdell}},\
  }\href {\doibase 10.1103/PhysRevA.83.063847} {\bibfield  {journal} {\bibinfo
  {journal} {Phys. Rev. A}\ }\textbf {\bibinfo {volume} {83}},\ \bibinfo
  {pages} {063847} (\bibinfo {year} {2011})}\BibitemShut {NoStop}%
\bibitem [{\citenamefont {Grudinin}\ \emph {et~al.}(2007)\citenamefont
  {Grudinin}, \citenamefont {Matsko},\ and\ \citenamefont
  {Maleki}}]{Grudinin2007}%
  \BibitemOpen
  \bibfield  {author} {\bibinfo {author} {\bibfnamefont {I.~S.}\ \bibnamefont
  {Grudinin}}, \bibinfo {author} {\bibfnamefont {A.~B.}\ \bibnamefont
  {Matsko}}, \ and\ \bibinfo {author} {\bibfnamefont {L.}~\bibnamefont
  {Maleki}},\ }\href {\doibase 10.1364/OE.15.003390} {\bibfield  {journal}
  {\bibinfo  {journal} {Opt. Express}\ }\textbf {\bibinfo {volume} {15}},\
  \bibinfo {pages} {3390} (\bibinfo {year} {2007})}\BibitemShut {NoStop}%
\bibitem [{\citenamefont {Kupriyanov}\ \emph {et~al.}(2005)\citenamefont
  {Kupriyanov}, \citenamefont {Mishina}, \citenamefont {Sokolov}, \citenamefont
  {Julsgaard},\ and\ \citenamefont {Polzik}}]{kupriyanov2005multimode}%
  \BibitemOpen
  \bibfield  {author} {\bibinfo {author} {\bibfnamefont {D.}~\bibnamefont
  {Kupriyanov}}, \bibinfo {author} {\bibfnamefont {O.}~\bibnamefont {Mishina}},
  \bibinfo {author} {\bibfnamefont {I.}~\bibnamefont {Sokolov}}, \bibinfo
  {author} {\bibfnamefont {B.}~\bibnamefont {Julsgaard}}, \ and\ \bibinfo
  {author} {\bibfnamefont {E.~S.}\ \bibnamefont {Polzik}},\ }\href@noop {}
  {\bibfield  {journal} {\bibinfo  {journal} {Physical Review A}\ }\textbf
  {\bibinfo {volume} {71}},\ \bibinfo {pages} {032348} (\bibinfo {year}
  {2005})}\BibitemShut {NoStop}%
\end{thebibliography}%
\onecolumngrid
\appendix

\section{Phase modulation}\label{app:pm}
Here we expand the spectral range of dispersion compensation by using additional phase modulation of the control field. We apply phase modulation applied only to the laser field acting on the $\ket{4}\leftrightarrow\ket{6}$ transition: \fla{
&\Omega_{\text{L}'}e^{-\text{i}(\omega_{46}-\Delta_{\text{L}'}) t}\rightarrow
\sum_{n=-n_\text{c}}^{n_\text{c}}\Omega^{(n)}_{\text{L}'}
e^{-\text{i}\omega_{n}t},
\label{Eq:phasemodulated}
}
\noindent
where $\omega_n=\omega_{46}-\Delta_{\text{L}'}+n\Delta_\text{SB}$, Rabi frequency of sidebands $\Omega^{(n)}_{\text{L}'}=\alpha^{\abs{n}}\Omega_{\text{L}'}$, $\alpha$ is the modulation depth, $2n_\text{c}+1$ is the overall number of spectral components.
If the atoms are out of two-photon resonance with the central frequency and the sidebands, the resultant dispersion is superimposed as though from several simultaneous $\Lambda$ schemes with different detunings. 
The phase-modulated field from Eq.~\eqref{Eq:phasemodulated} changes the dispersion interaction represented by the term $\beta_2(\omega)$.
At the same time, the first impedance-matching condition $\Re e[\beta_1(\omega)]=\kappa/2$ does not change.
Treating several control fields acting simultaneously  as independent $\Lambda$ schemes  is valid under the condition $\left|{\frac{\Omega^{*(n)}_{\text{L}'}(t)}{\Delta^{(n)}_{\text{L}'}+i \gamma_{46}}}\right|
\ll1$ of small off-resonant atomic excitation by modulated control field \cite{kupriyanov2005multimode}. 
By summing  contribution from all control field the dispersion compensation term  is modified to
\fla{    
&\beta_2(\omega)=-N_4\bar{P}_{4} |g_{56}|^2 
\frac{\sum_{n=-n_c}^{n_c}  \frac{|\Omega^{(n)}_{\text{L}'}|^2}{(\Delta^{(n)}_{\text{L}'}+i \gamma_{46})}
\left(\frac{1}{\delta^{(n)}_{\text{L}'}-\omega-i\gamma_{45}}+\frac{1}{\Delta^{(n)}_{\text{L}'}-i \gamma_{46}}\right) }
{(\Delta_{\text{L}'}+\delta -\omega-i\gamma_{56} )-\sum_{n=-n_c}^{n_c} \frac{\abs{\Omega^{(n)}_{\text{L}'}}^2}{(\delta^{(n)}_{\text{L}'}-\omega-i\gamma_{45})}},
\label{beta2_modul}
}
\noindent
where $\Delta^{(n)}_{\text{L}'}=\Delta_{\text{L}'}-n\Delta_\text{SB}$ and
$\delta^{(n)}_{\text{L}'}=\delta+n\Delta_\text{SB}$. This function allows to further increase bandwidth of the memory as it has more control parameters.


\section{Inhomogeneous broadening} \label{app:IB}
Inhomogeneous broadening is simulated in the continuous limit, where the sum over all atoms is replaced by an integral over their frequency with respect to central:
\fla{
\sum^{N}_{i=1} \rightarrow N\int^{\infty}_{-\infty} d\delta_{21} G(\delta_{21}),
}
where atoms have specific inhomogeneous broadening distribution $G(\delta_{21})$. We consider two cases (a) Lorentzian distribution 
$G(\delta_{21}) = \frac{\Delta_\text{in}}{\pi( \Delta^{2}_\text{in} + \delta^2_{21} ) }$ and
(b) uniform rectangular $G(\delta_{21}) = \frac{\text{rect}(\delta_{21}/\Delta_\text{in}) }{\Delta_\text{in}}$, where $\text{rect}(x)=\begin{cases} 1, \,\,|x| < 1/2,\\ \frac{1}{2}, \,|x| = 1/2,\\ 0,\,\, |x| > 1/2,\end{cases}$ is the rectangular function and $\Delta_\text{in}$ is the width of inhomogeneous broadening. For Lorentzian shape $\beta_1$
\fla{
\beta_1(\omega)&=\bar{P}_1\sum^{N_1}_{j=1}  \frac{ \left|g_{13}^j\right|^2 }{\gamma_{13}+\text{i}
(\Delta_\text{L}-\omega)
-\text{i}\frac{|\Omega_\text{L}|^2}{\delta_{21}^j-\omega-\text{i}\gamma_{12}}}\nonumber\\
&=N_1\bar{P}_1\left|g_{13}\right|^2\int \frac{G(\delta_{21})d\delta_{21} }{\gamma_{13}+\text{i}
(\Delta_\text{L}-\omega)
-\text{i}\frac{|\Omega_\text{L}|^2}{\delta_{21}-\omega-\text{i}\gamma_{12}}}\nonumber\\
&=\frac{N_1\bar{P}_1\left|g_{13}\right|^2\left(\omega+\text{i}(\gamma_{12} +\Delta_\text{in})\right)}{(\gamma_{12}+\Delta_\text{in}-i\omega)(\text{i}\gamma_{13}-\Delta_\text{L}+\omega)+\text{i}\abs{\Omega_\text{L}}^2}.
}
In derivation above we substituted sum with an integral, we used an approximation that coupling is independent of an atom $\left|g_{13}^j\right|^2\approx\left|g_{13}\right|^2$, and integrated over the Lorentzian inhomogeneous broadening.
In the case of uniform rectangular shape, the analytical solution has the more complicated form
\fla{
\beta_1(\omega)=&\bar{P}_1 \left|g_{13}\right|^2\sum^{N_1}_{j=1}  \frac{1 }{\gamma_{13}+\text{i}
(\Delta_\text{L}-\omega)
-\text{i}\frac{|\Omega_\text{L}|^2}{\delta_{21}^j-\omega-\text{i}\gamma_{12}}}\nonumber\\
=&\frac{N_1\bar{P}_1\left|g_{13}\right|^2}{\Delta_\text{in}}\int_{-\Delta_\text{in}/2}^{\Delta_\text{in}/2} \frac{d\delta_{21} }{\gamma_{13}+\text{i}
(\Delta_\text{L}-\omega)
-\text{i}\frac{|\Omega_\text{L}|^2}{\delta_{21}-\omega-\text{i}\gamma_{12}}}\nonumber\\
\approx &{N_1\bar{P}_1\left|g_{13}\right|^2}\bigg[\frac{1}{(\gamma_{13}+\text{i}(\Delta_\text{L}-\omega))}\nonumber\\
&+\frac{\text{i}\abs{\Omega_\text{L}}^2}{\Delta_\text{in}(\gamma_{13}+\text{i}(\Delta_\text{L}-\omega))^2}\log\left(\frac{(\gamma_{13}+\text{i}(\Delta_\text{L}-\omega))(\Delta_\text{in}/2-(\omega+\text{i}\gamma_{12}) )-\text{i}\abs{\Omega_{L}}^2}{(\gamma_{13}+\text{i}(\Delta_\text{L}-\omega))(-\Delta_\text{in}/2-(\omega+\text{i}\gamma_{12}) )-\text{i}\abs{\Omega_{L}}^2}\right)\bigg].
}
Similarly, for Gaussian inhomogeneous broadening $G(\delta_{21})=\frac{e^{-\delta^{2}_{21}/\left(2\Delta^2_{\text{in}}\right)}}{\sqrt{2\pi}\Delta_\text{in}}$,
\fla{
\beta_1(\omega) &= \int d \delta_{21} G(\delta_{21}) = \frac{1}{ (\gamma_{13}+i (\Delta_\text{L}- \omega) )}\nonumber \\
&-\frac{\sqrt{\frac{\pi }{2}} \Omega_{\text{L}}^2 \left(-1+\text{i}\cdot \text{erfi}\left(i\frac{(\gamma_{12} -i\omega) (\gamma_{13}  +i( \Delta_\text{L} -\omega) ) +\Omega_{L}^2}{\sqrt{2} \Delta_{\text{in}} (\gamma_{13}+i (\Delta_L-\omega ))}\right)\right) \exp \left(\frac{\left( (\gamma_{12}-i\omega) (\gamma_{13}+i (\Delta_\text{L}-\omega )) +\Omega_{\text{L}}^2 \right)^2}{2 \Delta_{\text{in}} ^2 (\gamma_{13}+i (\Delta_\text{L}-\omega ))^2}\right)}{\Delta_{\text{in}} (\gamma_{13}+i (\Delta_\text{L}- \omega) )^2}
}
These results are used to find final efficiencies of our memory.
\section{Inhomogeneous broadening for dispersion-compensation ensemble}
\label{Appx:InhomoDispComp}
\begin{figure}
  \includegraphics[width=0.49\columnwidth]{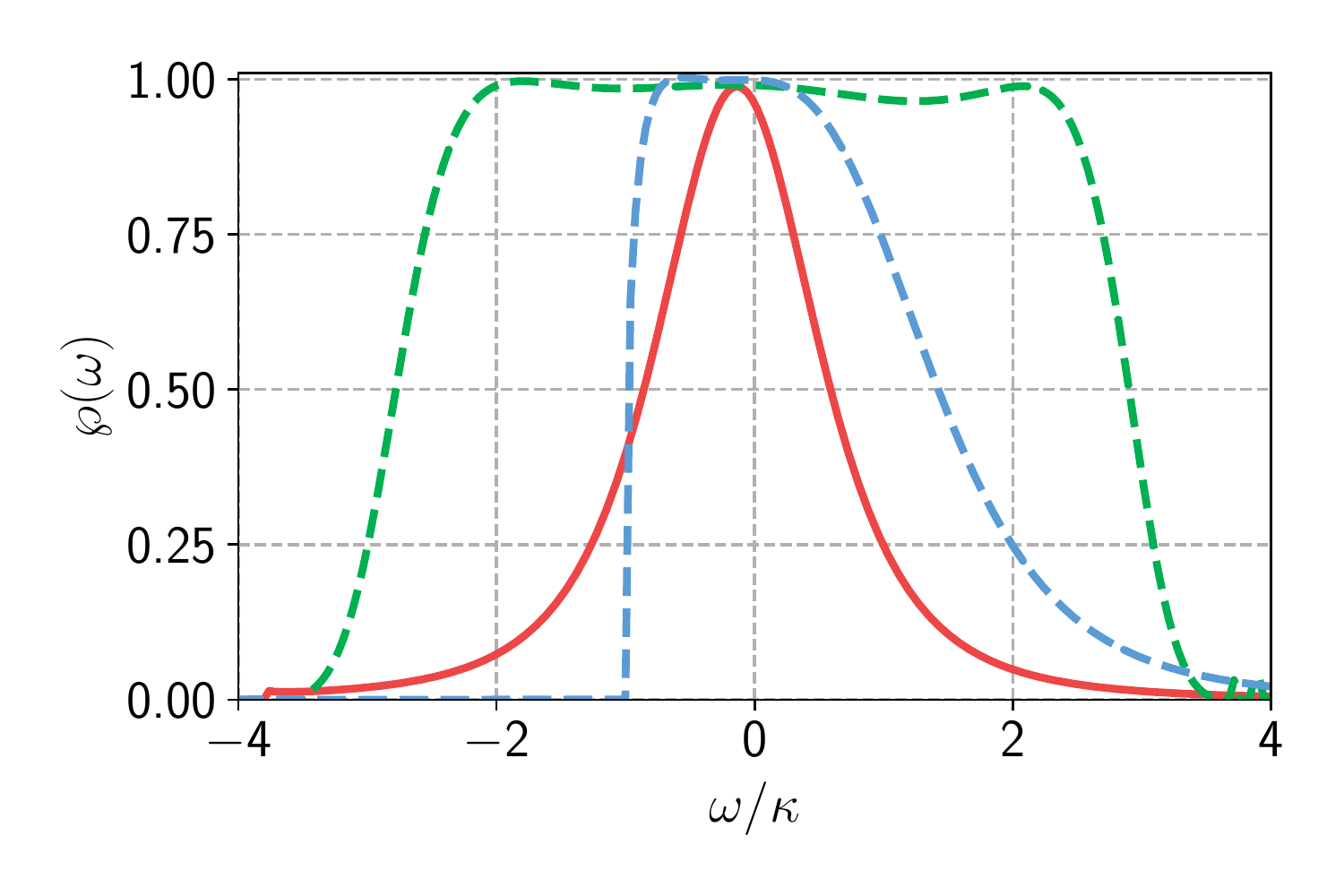}
\caption[Numerical simulation]{%
Spectral efficiency ~$\wp(\omega)$ vs scaled frequency~$\omega/\kappa$
for off-resonant Raman impedance-matched quantum memory (\textcolor{red1}{\textbf{red}} solid) and
with dispersion compensation (\textcolor{blue1}{\textbf{blue}} dashed) and with phase-modulated dispersion compensation (\textcolor{green1}{\textbf{green}} dashed) for uniform rectangular  inhomogeneous broadening.
The parameters are identical for the red and blue curves, 
normalized to $\kappa=1$, and are
$\gamma_{12}=\gamma_{45}=10^{-4}$,
$\gamma_{13}=\gamma_{46}=\gamma_{56}=0.1$, $\Omega_\text{L}=3$,
$\Omega_{\text{L}'}=13.55$,
$\Delta_\text{L}=150$,
$\Delta_{\text{L}'}=100$,
$\delta=-12.8$, $N_1 \bar{P}_1 \abs{g_{13}}^2=482$, $N_2 \bar{P}_4 \abs{g_{56}}^2=837$ and 
$\Delta_\text{in}=\Delta_\text{in,2}=14$.
The \textcolor{green1}{\textbf{green}} dashed line has some different parameters, which are $N_2 \bar{P}_4 \abs{g_{56}}^2=1700$,
$\delta=-12.7$, $\Delta_\text{SB}=21.1$, and $\alpha=0.87$.
}
\label{fig:inhomobroad45}
\end{figure}

Here we add inhomogeneous broadening to $\beta_2(\omega)$ (Eq.~\eqref{eq:beta2}) and integrate in the same manner as in Appendix \ref{app:IB}. For Lorentzian, Gaussian and uniform rectangular shapes,
\fla{
\beta_2 (\omega) =N_2\bar{P}_4|g_{56}|^2 
\begin{cases}
& \int  \frac{\Delta_{\text{in}}}{\pi \left( \Delta^{2}_{\text{in}} + (\delta' - \delta )^2 \right)}
\frac{\Omega_{\text{L}'}^2\left(\frac{1}{\Delta_{\text{L}'}+\text{i} \gamma_{46}}-\frac{\omega-\delta'+\text{i}\gamma_{45}}{\Delta_{\text{L}'}^2+\gamma_{46}^2}\right)}
{\Omega_{\text{L}'}^2+(\Delta_{\text{L}'}+\delta' -\omega-\text{i}\gamma_{56})(\omega-\delta'+\text{i} \gamma_{45})} \text{d}\delta' \,\,\quad \text{Lorentz}\\
& \int  \frac{ e^{-\frac{\delta'^{2}}{2\Delta_{\text{in}} }}   }{\sqrt{2\pi} \Delta_{\text{in}}}
\frac{\Omega_{\text{L}'}^2\left(\frac{1}{\Delta_{\text{L}'}+\text{i} \gamma_{46}}-\frac{\omega-\delta'+\text{i}\gamma_{45}}{\Delta_{\text{L}'}^2+\gamma_{46}^2}\right)}
{\Omega_{\text{L}'}^2+(\Delta_{\text{L}'}+\delta' -\omega-\text{i}\gamma_{56})(\omega-\delta'+\text{i} \gamma_{45})} \text{d}\delta' \,\,\qquad\quad\quad \text{Gaussian}\\
&  \int^{{\delta+\Delta_\text{in}/2}}_{\delta-\Delta_\text{in}/2}  \frac{\Omega_{\text{L}'}^2\left(\frac{1}{\Delta_{\text{L}'}+\text{i} \gamma_{46}}-\frac{\omega-\delta'+\text{i}\gamma_{45}}{\Delta_{\text{L}'}^2+\gamma_{46}^2}\right)}
{\Omega_{\text{L}'}^2+(\Delta_{\text{L}'}+\delta' -\omega-\text{i}\gamma_{56})(\omega-\delta'+\text{i} \gamma_{45})} \text{d}\delta'\,\,\qquad \qquad \text{uniform rectangular} \\
\end{cases}
}
In Figure \ref{fig:inhomobroad45} we find optimum parameters to maximize the width of our memory for uniform rectangular inhomogeneous broadening.
We conclude that rectangular broadening allows direct compensation of inhomogeneous broadening on levels $\ket{4}\leftrightarrow\ket{5}$ by increasing both an optical depth of dispersion compensation sub-ensemble and two-photon detuning. For the same set of experimental parameters the Lorentzian or Gaussian inhomogeneous broadening could not be compensated via free parameters. We attribute this to the fact that Lorentzian (Gaussian) broadening leads to pure dephasing. 
\end{document}